\documentclass[a4paper]{article}
\pdfoutput=1
\usepackage{a4wide}
\usepackage{booktabs}
\usepackage{graphicx,color}
\usepackage[usenames,dvipsnames]{xcolor}
\usepackage{amsmath,amssymb,amsfonts}
\usepackage{hyperref}
\usepackage{array}
\usepackage[small,bf]{caption}
\usepackage{cite}
\usepackage{bbm}

\newcommand{\gev}{\ensuremath{\,\mathrm{GeV}}}
\newcommand{\tev}{\ensuremath{\,\mathrm{TeV}}}

\newcommand{\mchi}{\ensuremath{m_{\tilde{\nu}_1}}}
\newcommand{\DM}{\ensuremath{{\tilde{\nu}_1}}}

\allowdisplaybreaks

\begin{document}

\begin{titlepage}

\vspace*{-15mm}
\begin{flushright}
NCTS-PH/1808
\end{flushright}
\vspace*{0.7cm}

\begin{center}

{\bf \Large Sneutrino Dark Matter via pseudoscalar $\boldsymbol{X}$-funnel \\[1ex]
meets Inverse Seesaw}
\\[8mm]

Jung Chang$^{\, a,b,}$\footnote{E-mail: \texttt{lovejesus99wwjd@gmail.com}},
Kingman Cheung$^{\, b,c,d,}$\footnote{E-mail: \texttt{cheung@phys.nthu.edu.tw}},
Hiroyuki Ishida$^{\, b,}$\footnote{E-mail: \texttt{hiroyuki403@cts.nthu.edu.tw}},
Chih-Ting Lu$^{\, c,}$\footnote{E-mail: \texttt{timluyu@hotmail.com}}, \\
Martin Spinrath$^{\, b,}$\footnote{E-mail: \texttt{martin.spinrath@cts.nthu.edu.tw}},
and Yue-Lin Sming Tsai$^{\, e,}$\footnote{E-mail: \texttt{smingtsai@gate.sinica.edu.tw}}
\\[1mm]
\end{center}
\vspace*{0.50cm}
\centerline{$^{a}$ \it Department of Physics, Chonnam National University,
300 Yongbong-dong,}
\centerline{\it Buk-gu, Gwangju, 500-757, Republic of Korea}
\vspace*{0.2cm}
\centerline{$^{b}$ \it Physics Division, National Center for Theoretical Sciences, Hsinchu 30013, Taiwan}
\vspace*{0.2cm}
\centerline{$^{c}$ \it Department of Physics, National Tsing Hua University, 
Hsinchu 30013, Taiwan}
\vspace*{0.2cm}
\centerline{$^{d}$ \it Division of Quantum Phases \& Devices, School of Physics, 
Konkuk University,}
\centerline{\it Seoul 143-701, Republic of Korea}
\vspace*{0.2cm}
\centerline{$^{e}$ \it Institute of Physics, Academia Sinica, Nangang, Taipei 11529, Taiwan}
\vspace*{1.20cm}

\begin{abstract}
\noindent
In this paper we study sneutrino dark matter in a recently proposed
supersymmetric electro\-weak-scale inverse seesaw model, 
in which the majority of the sneutrino dark matter particle is a mixture of the 
right-handed sneutrino $\tilde{N}^c$ and the singlet field $\tilde{S}$.
The scalar field $X$ responsible for the generation of neutrino masses can simultaneously
play a crucial role for sneutrino annihilation in the early Universe
via the pseudoscalar mediator $A_X$ into neutrinos.
We focus here on the dominant annihilation channels
and provide all the formulas together
with analytic estimates in order to identify the relevant parameters. 
Furthermore, we show that 
the direct detection scattering cross section is many orders of
magnitude below the current limits, and estimate 
the indirect detection annihilation rate,
which is only a few orders of magnitude below the current limits.
\end{abstract}

\end{titlepage}
\setcounter{footnote}{0}

\renewcommand{\thefootnote}{\arabic{footnote}}

\section{Introduction}
There are many well-established experimental hints 
for the existence of dark matter (DM) in the Universe, 
for instance, gravitational effects on visible matter, 
e.g., galactic rotational velocities, structure formation, etc.
However, the nature of DM is still unknown \cite{silk}.
So far all the terrestrial experiments looking for a direct
scattering signal with the DM particles fail to see anything. A large
portion of the mass range and interaction strength of the so-called
weakly-interacting massive particles (WIMP) are ruled out. The focus shifts
now to lighter mass ranges (sub-GeV) or other types
of DM like axions.

Another undeniable evidence of physics beyond the 
Standard Model (SM) is neutrino masses and oscillations \cite{osci}.
An interesting possibility is that DM may be deeply related to 
the generation of neutrino masses, for which we study an example here.
In a previous work, we have proposed a supersymmetric 
inverse seesaw (ISS) model~\cite{Chang:2017qgi}, in which 
the neutrino mass scale is related to the supersymmetry (SUSY) breaking scale.
This is in contrast to  many other ISS  models where 
the right-handed neutrino masses are forbidden and the smallness of the 
lepton number violating masses are put in by hand. 
The model that we proposed has a built-in $Z_3 \times Z_2$ symmetry, in
which the $Z_2$ factor is identical to the matter parity, which guarantees
the stability of the lightest supersymmetric particle (LSP) in our model
and all $R$-parity violating operators are automatically forbidden as we will explain in more detail later. 
Thus, the LSP can be a DM candidate. 
In this work, we work out explicitly the link between the neutrino mass generation
and the DM particle in our model.

In the conventional constrained Minimal Supersymmetric Standard Model (MSSM), 
the most popular LSP or DM candidate is the lightest neutralino, 
and the parameter space for viable DM fall into 
(i) co-annihilation region, (ii) focus-point region,
or (iii) $A$-funnel region. In all these three regions, fairly strong 
annihilation rates are needed for the LSP, so that it would not be
overproduced in the early Universe. In particular, the $A$-funnel region is
where the pseudoscalar mass is roughly twice of the DM mass 
such that the annihilation occurs very close to the resonance.
The DM candidate in our model shares a similar spirit as the
$A$-funnel in the MSSM as we will elaborate in subsequent sections. 

In the supersymmetric ISS model that we proposed 
\cite{Chang:2017qgi}, there are extra superfields $\hat{N}^c$, $\hat{S}$, 
and $\hat{X}$ in addition to the conventional superfields in the MSSM. 
Here $\hat{N}^c$ and $\hat{S}$ are 
lepton-like particles and their fermionic components mix with the SM 
neutrinos after symmetry breaking. The neutrino mass matrix exhibits an 
ISS-like structure and thus, light neutrino masses can be generated with
TeV-scale mass parameters of the model.
The vacuum expectation value (vev) of $X$ for that matter generates
the lepton number breaking components of the Majorana neutrino mass matrix. 
In this model the lightest sneutrino can be the DM candidate, which
is a scalar particle in contrast to the usual neutralino DM. 
The purely left-handed sneutrino DM in the MSSM 
faces two problems because of its sizeable couplings with the $Z$ boson:
i) the relic density is too small~\cite{Hagelin:1984wv,Ibanez:1983kw} 
and/or ii) a large direct detection cross section~\cite{Falk:1994es}.  
Some remedies exist in the literature where the sneutrino is a linear
combination of left-handed and right-handed components, 
e.g.,~\cite{ArkaniHamed:2000bq,Hooper:2004dc,Arina:2007tm,Arina:2008bb,Choi:2013fva}. 
There are even models where the lightest sneutrino is
purely right-handed and thus 
a viable DM candidate~\cite{Asaka:2005cn,Asaka:2006fs,
McDonald:2006if,Page:2007sh,Lee:2007mt,Cerdeno:2008ep,DelleRose:2017ukx}.
In our setup the lightest sneutrino is mostly a mixture of scalar components
of $\hat{N}$ and $\hat{S}$ but it contains a tiny fraction of left-handed sneutrinos as well.
We assume that the sneutrinos are produced thermally in the early Universe
and annihilate via the $H_X$- or $A_X$-funnel, where $H_X$($A_X$)
is the real (pseudo)scalar component of the complex $X$-scalar. We will show
that only the $A_X$-funnel is sufficiently efficient to get the right relic density.

The organization of the work is as follows. In the next section, we discuss
the stability of the lightest sneutrino as the DM using symmetry arguments
and compare our model to other supersymmetric models in the literature. 
In Sec.~\ref{sec:Masses}, 
we present formulas for the mass spectra for sneutrinos, neutralinos, and Higgs bosons 
and in Sec.~\ref{sec:DM}, we calculate the relic density and discuss estimates
for direct and indirect detection experiments.
Finally, we summarize and conclude in Sec.~\ref{sec:Summary}.

\section{Stability of the Lightest Supersymmetric Particle}
\label{sec:DMparity}

\begin{table}
\centering
\begin{tabular}{lcccccccccc}
\toprule
Superfield & $\hat{Q}_i$ & $\hat{U}_i^c$ & $\hat{E}_i^c$ & $\hat{L}_i$ & $\hat{D}_i^c$ & $\hat{H}_u$ & $\hat{H}_d$ & $\hat{N}^c_\alpha$ & $\hat{S}_\alpha$ & $\hat{X}$ \\
\midrule 
$Z_3$ charge $(q_3)$ & 1 & 1 & 1 & 0 & 0 & 1 & 2 & 2 & 1 & 1 \\
$Z_2$ charge $(q_2)$ & 1 & 1 & 1 & 1 & 1 & 0 & 0 & 1 & 1 & 0 \\
\bottomrule
\end{tabular}
\caption{\label{tab:Model}
Superfield content of the model and charge assignment under the additional
discrete $Z_6 = Z_3 \times Z_2$ symmetry. The new superfields compared to the
MSSM, $\hat{N}^c$, $\hat{S}$ and $\hat{X}$,
are singlets under the Standard Model gauge group. The indices $i = 1$, 2, 3
and $\alpha = 1$, 2 are generation indices.
}
\end{table}

The stability of the LSP is in many cases guaranteed by a (discrete) symmetry. 
In the MSSM, this is usually either $R$-parity, $P_R = (-1)^{3 (B-L) + 2s}$~\cite{Farrar:1978xj}, 
or matter parity, $P_M = (-1)^{3 (B-L)}$~\cite{MatterParity}. Here
$B$ stands for baryon number, $L$ for lepton number, 
and $s$ for the spin of the particle.

In our model \cite{Chang:2017qgi} we cannot use any of the two. Lepton
number is explicitly broken by the $\hat{X}^3$ term in the superpotential.
Strangely, although we cannot define matter or $R$-parity using the above
definitions, the $R$-parity violating operators of the MSSM are forbidden.
This is not an accident as we elucidate now. 
To understand the symmetries in our model better we can rewrite the $Z_6$
symmetry of the original model into an isomorphic $Z_3 \times Z_2$ symmetry, 
see Table~\ref{tab:Model}.

The $Z_2$ factor is identical to matter parity if we restrict ourselves to
the MSSM fields. Higgs fields are even under the $Z_2$ and matter fields
odd. This pattern is still true if we go to the full model. The right-handed
neutrinos, $\hat{N}$, and the lepton-like singlets, $\hat{S}$, are odd as they should
be since their fermionic components mix with the MSSM neutrinos. The scalar
component of $\hat{X}$, which is even under the $Z_2$ receives a vev and hence
behaves like the Higgs doublets. The $Z_2$ symmetry is hence nothing else than
a straight-forward extension of matter parity to our model. We just cannot
refer in its definition explicitly to any accidental symmetry of the SM.
It is also important to note that the $Z_2$ remains unbroken after symmetry breaking.

To understand the properties of DM it is convenient to include spin.
Now $R$- and matter parity are known to be equivalent since the product of
$(-1)^{2s}$ for the particles involved in any interaction vertex in
a theory that conserves angular momentum is always equal to $+1$. This is
as well true in our model so that we arrive at what we call DM parity
\begin{equation}
 P_{DM} = (-1)^{q_2 + 2 \, s} \;,
\end{equation}
where $q_2$ is the charge of the corresponding field under the $Z_2$ of
Table~\ref{tab:Model}. Sneutrinos with $q_2 = 1$ and $2 s = 0$
and neutralinos with $q_2 = 0$ and $2 s = 1$ are odd under DM
parity, $P_{DM} = -1$. Hence, these two particles are potential DM
candidates in our model%
\footnote{The gravitino with $q_2 =0$ and $2s = 3$ is a potential DM candidate as well,
which we nevertheless do not discuss here any further.}.

We want to comment here briefly on other models in the literature.
Of course, there is a mountain of papers on (SUSY) DM models
and it is clearly beyond the scope of our paper to attempt to give a complete review on that.
Therefore, we will only comment on supersymmetric inverse seesaw
models which discuss DM as well. Most models which we found
\cite{Arina:2008bb, BhupalDev:2012ru, Banerjee:2013fga, Guo:2013sna,
Ghosh:2014pwa, Cao:2017cjf}
simply extend (implicitly) $R$-parity treating the additional
lepton-like fields like MSSM matter fields without any discussion on
how this relates explicitly to the lepton number. This can be done
without any obvious harm since a neutrino Majorana mass term breaks
lepton number by two units and hence does not break $R$-parity explicitly.
In their works, sneutrino DM annihilates either via the resonance
of a $s$-channel Higgs mediator or through large annihilation into $WW, ZZ$, and $hh$.
In Ref.~\cite{Gogoladze:2014vea} the authors introduced a field, $\hat{S}$,
whose vev breaks lepton number by two units similar to our
$\hat{X}$ field. However, they did not specify any UV completion or
symmetries and instead referred to some previous works.
In Refs.~\cite{Kang:2011wb, Chen:2015yuz} they sketched a path towards
a UV completion introducing a non-renormalisable operator and
a symmetry breaking field $\hat{S}$. They also discussed some tentative
UV completions but did not fully specify the symmetries, field content
and symmetry breaking potentials.
In their approach, the sneutrino can be the asymmetric DM
and is favored to be near the electroweak scale.

In Ref.~\cite{DelleRose:2017ukx,Khalil:2011tb, An:2011uq, Borah:2012bb, Abdallah:2017gde}
the authors were more explicit (or refer to previous explicit works)
and broke a gauged $U(1)_{B-L}$ symmetry spontaneously. 
In \cite{An:2011uq, Borah:2012bb} to be more precise they started with 
a left-right symmetric setup inspired by a grand unified theory 
which contains a $U(1)_{B-L}$ factor in the breaking chain. 
The $R$-parity in this cases usually survives as a discrete subgroup ensuring DM stability.
Nevertheless, in Ref.~\cite{Borah:2012bb} $R$-parity was broken such that DM can decay.
In Ref.~\cite{DeRomeri:2012qd, Frank:2017ohg} they introduced an additional
$U(1)_{B-L} \times U(1)_R$ symmetry and assumed matter parity on top
to guarantee the stability of DM after symmetry breaking
and avoiding additional unwanted terms.

In this work, we will focus on thermally produced sneutrino DM
which co-annihilates through what we dubbed the $A_X$-funnel. In our setup 
the complex $X$-field hence plays two roles. First of all, the vev
of its CP-even component breaks
lepton number by two units generating neutrino masses via an inverse
seesaw mechanism. Secondly, its CP-odd scalar component plays a crucial role
to get the right thermal relic density.

In so far our paper differs from previous SUSY inverse seesaw
models which did not discuss explicitly the origin of lepton-number violation.
Most similar to our work we found are \cite{DelleRose:2017ukx,An:2011uq, Abdallah:2017gde},
where a heavy $B-L$ breaking Higgs or $Z'$ resonance plays an important role
for the DM annihilation cross section. However, the final state
contains usually more SM particles compared to our model, which makes the
phenomenology very different from ours.

\section{Masses of Sneutrinos, Neutralinos, and Higgs Fields}
\label{sec:Masses}

Before we discuss in detail DM phenomenology we first briefly
discuss the spectrum of the relevant fields. The approximate formulas, which
we present in this section, help a great deal in identifying the relevant
parameter space. 

For the convenience of the reader we recapitulate the superpotential of our
model, c.f.~\cite{Chang:2017qgi},
\begin{equation}
  \mathcal{W} = \mathcal{W}_{\text{MSSM}} + \mathcal{W}_\nu \;,
\end{equation}
where
\begin{align}
\mathcal{W}_{\text{MSSM}} &= Y_u \, \hat{Q} \hat{H}_u \hat{U}^c - Y_d \, \hat{Q} \hat{H}_d \hat{D}^c - Y_e \, \hat{L} \hat{H}_d \hat{E}^c  + \mu_H \hat{H}_u \hat{H}_d \;, \\
\mathcal{W}_\nu       &= Y_\nu \, \hat{L} \hat{H}_u \hat{N}^c + \mu_{NS} \, \hat{N}^c \hat{S} + \frac{\lambda}{2} \, \hat{X} \, \hat{S}^2 + \frac{\kappa}{3} \, \hat{X}^3 \;.
\end{align}
Beyond the MSSM Yukawa couplings and $\mu_H$-term we have three new Yukawa
couplings $Y_\nu$, $\lambda$ and $\kappa$ and a new mass parameter $\mu_{NS}$.

Similarly the soft SUSY breaking terms can be grouped into an ordinary MSSM part
and additional terms
\begin{equation}
 - {\cal L}_{\rm soft} = - {\cal L}_{\rm soft, MSSM} - {\cal L}_{\rm soft, \nu} \;,
\end{equation}
where 
\begin{align}
- {\cal L}_{\rm soft, MSSM}  &= \frac{1}{2} M_1 \tilde{B} \tilde{B} 
              + \frac{1}{2} M_2 \tilde{W} \tilde{W}
              + \frac{1}{2} M_3 \tilde{g} \tilde{g} \nonumber \\
  &+ M_{\tilde{Q}}^2 \tilde{Q}^\dagger \tilde{Q} 
     +  M_{\tilde{U^c}}^2 \tilde{U^c}^\dagger \tilde{U_c} 
     +  M_{\tilde{D^c}}^2 \tilde{D^c}^\dagger \tilde{D_c}
     +  M_{\tilde{L}}^2 \tilde{L}^\dagger \tilde{L} 
     +  M_{\tilde{E^c}}^2 \tilde{E^c}^\dagger \tilde{E^c}\nonumber \\
  &  + M_{H_u}^2 {H_u}^\dagger H_u + M_{H_d}^2 {H_d}^\dagger H_d + (b_H H_u H_d +  \text{H.c.}) \nonumber \\
  &  + \left ( A_u \tilde{Q} H_u \tilde{U}^c - A_d \tilde{Q} H_d \tilde{D}^c 
     - A_e \tilde{L} H_d \tilde{E}^c + \text{H.c.} \right)  \;, \\
- {\cal L}_{\rm soft, \nu}  &= M_{\tilde{N^c}}^2 \tilde{N}^{c \dagger} \tilde{N}^c
     +  M_{\tilde{S}}^2 \tilde{S}^\dagger \tilde{S}
     + M_{X}^2 X^\dagger X
     + (b_{NS} \tilde{N}^c \tilde{S} + \text{H.c.}) \nonumber \\
  &  + \left ( A_\nu \tilde{L} H_u \tilde{N}^c 
     + \frac{1}{2} A_{\lambda} X {\tilde{S}}^2
     + \frac{1}{3} A_{\kappa} X^3 + \text{H.c.} \right )\;. 
\end{align}
For each new Yukawa coupling we have introduced a new trilinear coupling,
$A_\nu$, $A_{\lambda}$ and $A_{\kappa}$. Furthermore, we have introduced
$b_{NS}$ which corresponds to $\mu_{NS}$ in the superpotential and mass parameters
for the new scalar fields, $M_{\tilde{N^c}}^2$, $M_{\tilde{S}}^2$ and $M_{X}^2$.
For better readability we have suppressed here any flavor and gauge indices which
can be easily reconciled from Table~\ref{tab:Model}.

\subsection{Sneutrinos}
\label{sec:SneutrinoMasses}

We begin our discussion with the scalar partners of the neutrinos.
Compared to the MSSM our model contains many more sneutrinos.
As we had discussed in our previous paper \cite{Chang:2017qgi}
the leading-order expression for the sneutrino masses reads
\begin{equation} 
m^2_{\tilde{\nu}^R} \approx m^2_{\tilde{\nu}^I} \approx \left( 
\begin{array}{ccc}
\Re (M_{\tilde{L}}^2 ) + \tfrac{1}{2} M_Z^2 \cos (2 \beta) & 0 & 0 \\ 
0 & \Re(M_{\tilde{N}^c}^2  + \mu_{NS} \mu_{NS}^\dagger) & \Re(b_{NS})\\ 
0  & \Re(b_{NS}^T) & \Re(M_{\tilde{S}}^2  + \mu_{NS}^\dagger \mu_{NS}) \end{array} 
\right)\;.
\label{eq:SneutrinoMasses}
\end{equation}
In our numerical results later on, we use GUT-scale boundary conditions
where all bilinear sfermion mass parameters are set to $m_0$ at the GUT scale,
c.f.\ next section.
For simplicity and clarity we focus for the moment on the case of
one generation of left-handed, right-handed and singlet scalars each.
At the GUT scale the approximate mass matrix then further simplifies to
\begin{equation} 
m^2_{\tilde{\nu}^R} \approx m^2_{\tilde{\nu}^I} \approx \left( 
\begin{array}{ccc}
m_0^2 + \tfrac{1}{2} M_Z^2 \cos (2 \beta) & 0 & 0 \\ 
0 & m_0^2  + \mu_{NS}^2  & m_0^2\\ 
0  & m_0^2 & m_0^2  + \mu_{NS}^2 \end{array} 
\right) \;.
\label{eq:SneutrinoMassesSimple}
\end{equation} 
For $\mu_{NS} = 0$ one sneutrino is strictly massless
which might lead to phenomenological issues. 
Nevertheless, we assume $\mu_{NS}$ to be of the order of the electroweak scale, 
so we will not discuss this case any further.

To get an estimate of the sneutrino masses at the electroweak scale we discuss now
the renormalization group (RG) corrections to the leading-order expression in Eq.~\eqref{eq:SneutrinoMasses}.
For the one-loop $\beta$-functions we use the results calculated by \texttt{SARAH}~\cite{SARAH}
\begin{align}
\beta_{\mu_{NS}}^{(1)} 
& =  
2 {Y_{\nu}  Y_{\nu}^{\dagger}  \mu_{NS}}  + {\mu_{NS}  \lambda^*  \lambda} \;,\\ 
 \beta_{b_{NS}}^{(1)} 
& =  
2 {\mu_{NS}  \lambda^*  A_{\lambda}}  + 2 {Y_{\nu}  Y_{\nu}^{\dagger}  b_{NS}}  + 4 {A_{\nu}  Y_{\nu}^{\dagger}  \mu_{NS}}  + {b_{NS}  \lambda^*  \lambda} \;,\\ 
\beta_{M_{\tilde{L}}^2}^{(1)} 
& =  
-\frac{6}{5} g_{1}^{2} {\bf 1} |M_1|^2 - 6 g_{2}^{2} {\bf 1} |M_2|^2 +2 m_{H_d}^2 {Y_{e}^{\dagger}  Y_e} + 2 m_{H_u}^2 {Y_{\nu}^{\dagger}  Y_{\nu}} +2 {A_{e}^{\dagger}  A_e} +2 {A_{\nu}^{\dagger}  A_{\nu}} \nonumber  \\ 
&+ 
{M_{\tilde{L}}^2  Y_{e}^{\dagger}  Y_e}+{M_{\tilde{L}}^2  Y_{\nu}^{\dagger}  Y_{\nu}}+2 {Y_{e}^{\dagger}  M_{\tilde{E}}^2  Y_e} +{Y_{e}^{\dagger}  Y_e  M_{\tilde{L}}^2}+2 {Y_{\nu}^{\dagger}  M_{\tilde{N}^c}^2  Y_{\nu}} +{Y_{\nu}^{\dagger}  Y_{\nu}  M_{\tilde{L}}^2}\nonumber  \\ 
&- 
\frac{3}{5} g_1^2 {\bf 1} \Big(-2 \mbox{Tr}\Big({M_{\tilde{U}}^2}\Big)  - \mbox{Tr}\Big({M_{\tilde{L}}^2}\Big)  - m_{H_d}^2  + m_{H_u}^2 + \mbox{Tr}\Big({M_{\tilde{D}}^2}\Big) + \mbox{Tr}\Big({M_{\tilde{E}}^2}\Big) + \mbox{Tr}\Big({M_{\tilde{Q}}^2}\Big)\Big) \;,\\
\beta_{M_{\tilde{N}^c}^2}^{(1)} 
&= 
2 \Big(2 {A_{\nu}  A_{\nu}^{\dagger}}  + 2 m_{H_u}^2 {Y_{\nu}  Y_{\nu}^{\dagger}}  + 2 {Y_{\nu}  M_{\tilde{L}}^2  Y_{\nu}^{\dagger}}  + {M_{\tilde{N}^c}^2  Y_{\nu}  Y_{\nu}^{\dagger}} + {Y_{\nu}  Y_{\nu}^{\dagger}  M_{\tilde{N}^c}^2}\Big) \;,\\ 
\beta_{M_{\tilde{S}}^2}^{(1)} 
&= 
2 {A_{\lambda}^*  A_{\lambda}}  + 2 {\lambda^*  M_{\tilde{S}}^{2 *}  \lambda}  + 2 M_{X}^2 {\lambda^*  \lambda}  + {M_{\tilde{S}}^2  \lambda^*  \lambda} + {\lambda^*  \lambda  M_{\tilde{S}}^2} \;.
\end{align}
We assume that all trilinear couplings are equal to the product of $A_0$ 
and the corresponding Yukawa coupling. 
To get the expression in Eq.~\eqref{eq:SneutrinoMasses}
we had neglected all terms proportional to some powers of the small expansion
parameters $\epsilon$, c.f.~\cite{Chang:2017qgi}. 
In particular, if we neglected terms proportional to some powers of $Y_\nu$ and $\lambda$, 
we could simplify the $\beta$-functions as well to
\begin{align}
\beta_{\mu_{NS}}^{(1)} &=  \mathcal{O}(\epsilon^2) \;,\\ 
\beta_{b_{NS}}^{(1)} &= \mathcal{O}(\epsilon^2) \;,\\ 
\beta_{M_{\tilde{L}}^2}^{(1)} &= 
-\frac{6}{5} g_{1}^{2} {\bf 1} |M_1|^2 - 6 g_{2}^{2} {\bf 1} |M_2|^2 
+ 2 m_{H_d}^2 {Y_{e}^{\dagger}  Y_e} + 2  {A_{e}^{\dagger} A_e}
+{M_{\tilde{L}}^2  Y_{e}^{\dagger}  Y_e} + 2 {Y_{e}^{\dagger}  M_{\tilde{E}}^2  Y_e} + {Y_{e}^{\dagger}  Y_e  M_{\tilde{L}}^2} \nonumber  \\ 
&- \frac{3}{5} g_1^2 {\bf 1} \Big(-2 \mbox{Tr}\Big({M_{\tilde{U}}^2}\Big)  - \mbox{Tr}\Big({M_{\tilde{L}}^2}\Big)  - m_{H_d}^2  + m_{H_u}^2 + \mbox{Tr}\Big({M_{\tilde{D}}^2}\Big) + \mbox{Tr}\Big({M_{\tilde{E}}^2}\Big) + \mbox{Tr}\Big({M_{\tilde{Q}}^2}\Big)\Big) \nonumber\\
&+ \mathcal{O}(\epsilon^2)\;,\\
\beta_{M_{\tilde{N}^c}^2}^{(1)} &= \mathcal{O}(\epsilon^2) \;,\\ 
\beta_{M_{\tilde{S}}^2}^{(1)} &=  \mathcal{O}(\epsilon^2) \;.
\end{align}
This implies that the mass matrix for the right-handed sneutrinos and scalar singlets is very
well approximated by the high-scale boundary conditions. Only the left-handed sneutrino mass
parameters have sizeable one-loop running, which is positive (at the low scale $M_{\tilde{L}}^2 > m_0^2$).

The simplified expression for the sneutrino mass matrix in
Eq.~\eqref{eq:SneutrinoMassesSimple}
can be easily diagonalised with 
three eigenvalues $\mu_{NS}^2$, $\mu_{NS}^2 + m_0^2$, and
$M_{\tilde{L}}^2 + \tfrac{1}{2} M_Z^2 \cos (2 \beta)$. It is straightforward 
to write down the approximate eigenstates
\begin{align}
 \tilde{\nu}_{1,2} \approx \frac{1}{\sqrt{2}} \left( \tilde{N}^c_1 \mp \tilde{S}_1 \right) \text{ and }
 \tilde{\nu}_{3} \approx \tilde{L}_1 \;.\label{eq:DM-comp}
\end{align}
Since we know that the predominantly left-handed sneutrino DM 
has already been excluded by
direct detection experiments \cite{Falk:1994es,Arina:2007tm},
sneutrinos can only be DM if it is
mostly right-handed and/or scalar singlet, i.e.,
if $\mu_{NS}^2 \lesssim m_0^2 + M_Z^2$.
In fact, we expect that the second lightest sneutrino in each
generation is the left-handed one.
The sneutrino DM candidate is then, to a good approximation,
a maximally mixed superposition of the right-handed sneutrino and the scalar singlet.
Strictly speaking we have two DM candidates since the real and the
imaginary parts of the sneutrinos are very close in mass
and both can contribute to the DM density.
In fact, as we will see later on they both contribute to the relic density
and we have a two component DM. 

To understand under what conditions the real or imaginary component is
the lighter one and when both states can co-annihilate, we discuss now
the leading tree-level corrections to the masses of the lightest sneutrinos. 
Using ordinary perturbation theory it is straight-forward to find that
\begin{align}
 m_{\tilde{\nu}_1}^2 \approx \mu_{NS}^2 + \langle 1 | \delta M| 1 \rangle \;,\label{Eq:snutree}
\end{align}
where $\sqrt{2} \, |1\rangle = (0,1,-1)$ and $\delta M$ is
\begin{equation}
\delta M_R = \begin{pmatrix}
              \tfrac{1}{2} v_u^2 Y_\nu^2 & - \tfrac{Y_\nu}{\sqrt{2}} (v_d \mu_H - v_u A_0) & \tfrac{Y_\nu}{\sqrt{2}} v_u \mu_{NS} \\
              - \tfrac{Y_\nu}{\sqrt{2}} (v_d \mu_H - v_u A_0) & \tfrac{1}{2} v_u^2 Y_\nu^2 & \tfrac{\lambda}{\sqrt{2}} v_X \mu_{NS} \\
              \tfrac{Y_\nu}{\sqrt{2}} v_u \mu_{NS} & \tfrac{\lambda}{\sqrt{2}} v_X \mu_{NS} & \tfrac{1}{2} \lambda v_X (\sqrt{2} A_0 + (\kappa + \lambda ) v_X ) )
             \end{pmatrix}\,,
\end{equation}
for the CP-even sneutrinos and
\begin{equation}
\delta M_I = \begin{pmatrix}
              \tfrac{1}{2} v_u^2 Y_\nu^2 & - \tfrac{Y_\nu}{\sqrt{2}} (v_d \mu_H - v_u A_0) & \tfrac{Y_\nu}{\sqrt{2}} v_u \mu_{NS} \\
              - \tfrac{Y_\nu}{\sqrt{2}} (v_d \mu_H - v_u A_0) & \tfrac{1}{2} v_u^2 Y_\nu^2 & -\tfrac{\lambda}{\sqrt{2}} v_X \mu_{NS} \\
              \tfrac{Y_\nu}{\sqrt{2}} v_u \mu_{NS} & -\tfrac{\lambda}{\sqrt{2}} v_X \mu_{NS} & - \tfrac{1}{2} \lambda v_X (\sqrt{2} A_0 + (\kappa - \lambda) v_X ) )
             \end{pmatrix}\,,
\end{equation}
for the CP-odd sneutrinos. Note that we do not distinguish here between
the different ISS cases \cite{Chang:2017qgi}, which would further simplify
the expressions if $Y_\nu \ll \lambda$
or $\lambda \ll Y_\nu$.

The corrections of the lightest sneutrino masses read
\begin{align}
 \langle 1 | \delta M_R | 1 \rangle &= \frac{Y_\nu^2}{4} v_u^2 + \frac{1}{4} \lambda \, v_X \left( \sqrt{2} \, A_0 - 2\sqrt{2} \, \mu_{NS} + (\kappa + \lambda) \, v_X \right) \;, \label{Eq:Rsnucorr}\\
 \langle 1 | \delta M_I | 1 \rangle &= \frac{Y_\nu^2}{4} v_u^2 -  \frac{1}{4} \lambda \, v_X \left( \sqrt{2} \, A_0 - 2 \sqrt{2} \, \mu_{NS} + (\kappa - \lambda) \, v_X \right) \;, \label{Eq:Isnucorr}\\
 &\Rightarrow m_{\tilde{\nu}^R_1}^2 - m_{\tilde{\nu}^I_1}^2 \approx \frac{1}{2} \, \lambda \, v_X \left( \sqrt{2} \, A_0 - 2 \sqrt{2} \mu_{NS} + \kappa \, v_X \right) \;. \label{Eq:Diffcorr}
\end{align}
The mass splitting between the two lightest sneutrinos is
of the order of $\lambda \, M_{\text{SUSY}}^2$ and vanishes
for $v_X = 0$ as expected.

\subsection{Neutralinos}

\begin{figure}
 \centering
 \includegraphics[scale=0.5]{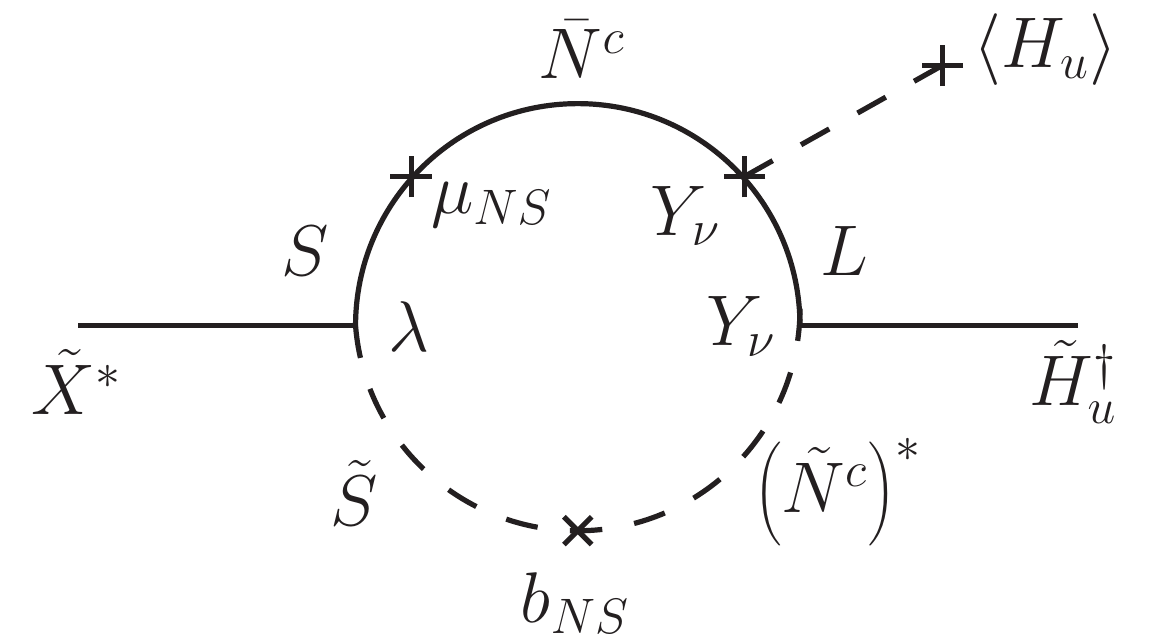}~~
 \includegraphics[scale=0.5]{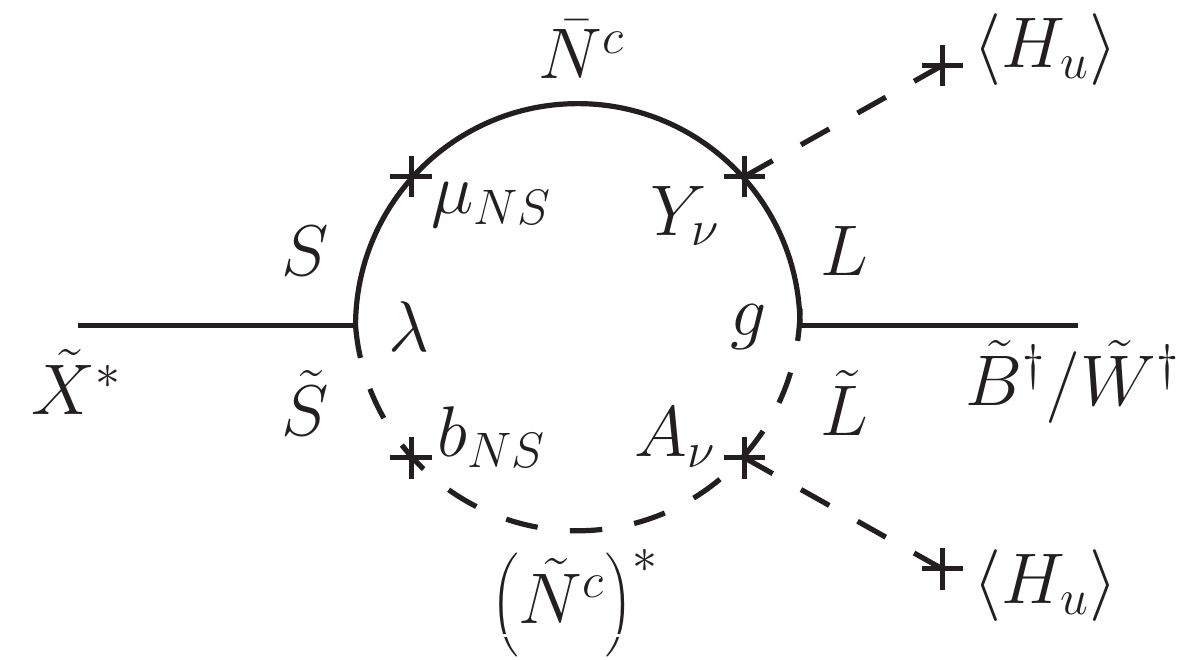}
 \caption{Typical one-loop diagrams inducing a mixing between the Xino and
  the MSSM neutralinos.
 \label{fig:XinoMixing}}
\end{figure}

Technically speaking the Xino, $\tilde{X}$, is a neutralino since it has the
same quantum numbers as the other MSSM neutralinos and will in general
mix with them. On tree level the neutralino mass
 matrix in the basis
\( \left(\tilde{B}, \tilde{W}^0, \tilde{H}_d^0, \tilde{H}_u^0, \tilde{X}\right) \)
has the structure
\begin{equation} 
M_{\tilde{\chi}^0} = \left( 
\begin{array}{ccccc}
M_1 &0 &-\frac{1}{2} g_1 v_d  &\frac{1}{2} g_1 v_u  &0\\ 
0 &M_2 &\frac{1}{2} g_2 v_d  &-\frac{1}{2} g_2 v_u  &0\\ 
-\frac{1}{2} g_1 v_d  &\frac{1}{2} g_2 v_d  &0 &- \mu_H  &0\\ 
\frac{1}{2} g_1 v_u  &-\frac{1}{2} g_2 v_u  &- \mu_H  &0 &0\\ 
0 &0 &0 &0 &\sqrt{2} \kappa v_{X} \end{array} 
\right) \;,
\end{equation}
which we have confirmed independently using \texttt{SARAH}~\cite{SARAH}.

Since the mixing of the Xino with the other neutralinos is absent
on tree-level we expect it to be small. In Fig.~\ref{fig:XinoMixing}
we show typical one-loop diagrams
which generate such mixings. Since all mass parameters are assumed to
be of the order of the electroweak scale it is easy to estimate the
size of the mixing terms
\begin{align}
 (\delta M_{\tilde{\chi}^0})_{i5} \sim \frac{1}{16 \pi^2} \lambda \, Y_\nu^2 \, 
\times \mathcal{O}(\text{TeV}) \;,
\end{align}
where we have assumed that $A_\nu \sim Y_\nu \times
 \mathcal{O}(\text{TeV})$.
Remembering that $m_\nu \sim  Y_\nu \lambda Y_\nu^T \times
\mathcal{O}(\text{TeV})$
in our model \cite{Chang:2017qgi}
we see that the mixing terms are of the size of the neutrino masses suppressed
by one additional loop factor which can be safely neglected. This is confirmed
as well by our numerical results.

Let us comment briefly on the Xino properties. 
The Xino could also be the LSP depending on the parameter choice.  
However, its dominant annihilation channel into $A_X H_X$ has in general
a too large cross section to give the right relic density 
because of the $\mathcal{O}(1)$ coupling $\kappa$. 
Only when the phase space for this process closes the annihilation cross section
could be sufficiently suppressed but we do not discuss this possibility any further.

The other four neutralinos are just the ordinary, well-known MSSM neutralinos
and we do not discuss them any further here.

\subsection{Higgs Bosons}
\label{sec:HiggsMasses}

The third sector, the Higgs sector, differs 
from that of the ordinary MSSM, by having two additional scalars. 
We decompose the scalar component of the superfield $\hat{X}$ as
\begin{equation}
 X = \frac{1}{\sqrt{2}} \left( v_X + \phi_X + \text{i} \, \sigma_X \right) \;.
\end{equation}
Note that we have used here a different normalisation convention for the vev
compared to our original paper \cite{Chang:2017qgi} in order to 
directly use the results
from \texttt{SARAH}~\cite{SARAH} without tedious checks for factors 
of $\sqrt{2}$. For
all the results in this paper we have switched to 
the \texttt{SARAH} conventions.

The scalar mass matrix then reads on tree-level in the basis
\( \left(\phi_{d}, \phi_{u}, \phi_{X}\right) \): 
\begin{equation} 
m^2_{h} = \left( 
\begin{array}{ccc}
m_{\phi_{d}\phi_{d}}^2 &-\frac{1}{4} \Big(g_{1}^{2} + g_{2}^{2}\Big)v_d v_u  - {\Re\Big(b_H\Big)}  &0\\ 
-\frac{1}{4} \Big(g_{1}^{2} + g_{2}^{2}\Big)v_d v_u  - {\Re\Big(b_H\Big)}  &m_{\phi_{u}\phi_{u}}^2 &0\\ 
0 &0 &m_{\phi_{X}\phi_{X}}^2\end{array} 
\right) \;,
 \end{equation} 
where
\begin{align} 
m_{\phi_{d}\phi_{d}}^2 &= \frac{1}{8} \Big(g_{1}^{2} + g_{2}^{2}\Big)\Big(3 v_{d}^{2}  - v_{u}^{2} \Big) + M_{H_d}^2 + |\mu_H|^2\,,\\ 
m_{\phi_{u}\phi_{u}}^2 &= -\frac{1}{8} \Big(g_{1}^{2} + g_{2}^{2}\Big)\Big(-3 v_{u}^{2}  + v_{d}^{2}\Big) + M_{H_u}^2 + |\mu_H|^2\,,\\ 
m_{\phi_{X}\phi_{X}}^2 &= v_{X} \Big(3 v_{X} |\kappa|^2  + \sqrt{2} {\Re\Big(A_{\kappa}\Big)} \Big) + M_{X}^2 \;.
\end{align}
The fields $\phi_d$ and $\phi_u$ are the CP-even and electrically neutral
components of the MSSM Higgs doublets $H_d$ and $H_u$, respectively. 
Similarly the fields $\sigma_d$ and $\sigma_u$ are their CP-odd components
and the mass matrix for the CP-odd Higgs-like scalars reads
on tree-level in the basis \( \left(\sigma_{d}, \sigma_{u}, \sigma_{X}\right) \):
\begin{equation} 
m^2_{A^0} = \left( 
\begin{array}{ccc}
m_{\sigma_{d}\sigma_{d}}^2 &{\Re\Big(b_H\Big)} &0\\ 
{\Re\Big(b_H\Big)} &m_{\sigma_{u}\sigma_{u}}^2 &0\\ 
0 &0 &m_{\sigma_{X}\sigma_{X}}^2\end{array} 
\right) \;,
 \end{equation} 
 where 
\begin{align} 
m_{\sigma_{d}\sigma_{d}}^2 
&= 
\frac{1}{8} \Big(g_{1}^{2} + g_{2}^{2}\Big)\Big(- v_{u}^{2}  + v_{d}^{2}\Big) + M_{H_d}^2 + |\mu_H|^2\,,\\ 
m_{\sigma_{u}\sigma_{u}}^2 
&= 
-\frac{1}{8} \Big(g_{1}^{2} + g_{2}^{2}\Big)\Big(- v_{u}^{2}  + v_{d}^{2}\Big) + M_{H_u}^2 + |\mu_H|^2\,,\\ 
m_{\sigma_{X}\sigma_{X}}^2 
&= 
v_{X} \Big(- \sqrt{2} {\Re\Big(A_{\kappa}\Big)}  + v_{X} |\kappa|^2 \Big) + M_{X}^2\,,
\end{align} 
and we have neglected gauge fixing contributions.

Similar to the Xino and the MSSM neutralinos, our new scalars do
not mix with the MSSM Higgs fields
at tree-level and the loop-level mixing is negligibly small
using similar arguments as for the Xino case. Hence, we label the new
scalars as follows $\phi_X \equiv H_X$ and $\sigma_X \equiv A_X$,
which are mass and symmetry eigenstates simultaneously to a very good
approximation.
Note that $M_{X}^2$ is fixed by the tadpole condition
\begin{align}
 \frac{\partial V}{\partial \phi_{X}} 
 &= 
 \frac{v_{X}^{2}}{\sqrt{2}}  {\Re\Big(A_{\kappa}\Big)}  + M_{X}^2 v_{X}  + v_{X}^{3} |\kappa|^2 = 0 \\
 &
 \Leftrightarrow M_X^2 = - \frac{v_X}{\sqrt{2}} {\Re\Big(A_{\kappa}\Big)}  - v_{X}^2 |\kappa|^2 \;.
\end{align}
We can use this in the formulas for the scalar masses
\begin{align}
 m_{H_X}^2 &=  2 \,  |\kappa|^2 v_{X}^2  + \frac{v_X}{\sqrt{2}} {\Re\Big(A_{\kappa}\Big)} \;, \\
 m_{A_X}^2 &=  - \frac{3 \, v_X}{\sqrt{2}} {\Re\Big(A_{\kappa}\Big)} \;.\label{eq:XHiggsmass}
\end{align}

\begin{figure}
 \centering
 \includegraphics[scale=1.2]{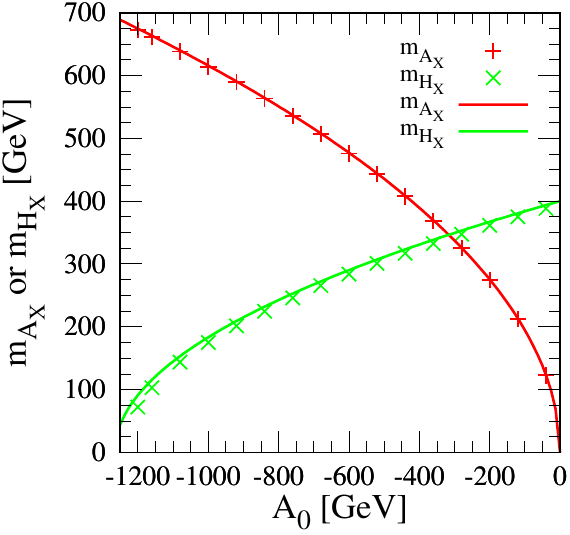}
 \caption{
 Comparison of the analytic results (straight lines) for the masses
 of $H_X$ and $A_X$ (Eqs.~\eqref{eq:HXMassEstimate} and \eqref{eq:AXMassEstimate})
 with numerical \texttt{SPheno} results (crosses) as a function of $A_0$. 
 Here, we take $\kappa = 0.4$ and $\sqrt{2} v_X=10^3~\gev$. 
 \label{Fig:Higgsmasses}
 }
\end{figure}

Before we quantify this equations further, we have to comment
on the running of $A_\kappa$ which is significant. The relevant one-loop
$\beta$-functions are given by
\begin{align}
 \beta_{\kappa} &= 6 \kappa |\kappa|^2  + \frac{3}{2} \kappa \text{Tr}(\lambda \lambda^*)  \approx 6 \kappa |\kappa|^2  \;, \\
 \beta_{A_{\kappa}} &= 18 |\kappa|^2 A_\kappa + 3 \, \kappa \, \text{Tr}(\lambda^* A_\lambda) + \frac{3}{2} A_\kappa \text{Tr}(\lambda \lambda^*) \approx 18 |\kappa|^2 A_\kappa \;.
\end{align}
Neglecting the small terms proportional to some powers of
$\lambda$ we can solve this set of coupled differential equations
analytically and find
\begin{align}
 \kappa(\mu) 
 &= 
 \frac{\kappa_0}{\sqrt{1 + \frac{3 \, \kappa_0^2}{4 \pi^2} \log (M_{\text{SUSY}}/\mu) }} \;,\\
 A_\kappa(\mu) 
 &= 
 \kappa_0 A_0 \left( \frac{(1 +\tfrac{3 \, \kappa_0^2}{4 \pi^2} \log (M_{\text{SUSY}}/M_{\text{GUT}}) )^2}{(1 + \tfrac{3 \, \kappa_0^2}{4 \pi^2} \log (M_{\text{SUSY}}/\mu) )^3} \right)^{\frac{1}{2}} \;,
\end{align}
where $\kappa_0 = \kappa(M_{\text{SUSY}})$. At the low scale where
we want to evaluate the scalar masses
\begin{align}
 A_\kappa(M_{\text{SUSY}}) 
 &= 
 \kappa_0 A_0 \left( 1 + \frac{3 \, \kappa_0^2}{4\pi^2} \log (M_{\text{SUSY}}/M_{\text{GUT}}) \right) \approx \kappa_0 A_0 \left( 1 - 2.3 \, \kappa_0^2 \right) \;,
\end{align}
where we have used $M_{\text{GUT}} \approx 2 \times 10^{16}$~GeV and
$M_{\text{SUSY}} \approx 10^{3}$~GeV.
At the SUSY scale we find for the RG corrected scalar masses
\begin{align}
\label{eq:HXMassEstimate}
 m_{H_X}^2 
 &\approx  
 2 \,  \kappa_0^2 v_{X}^2  + \frac{v_X}{\sqrt{2}} \kappa_0 A_0 \left( 1 - 2.3 \, \kappa_0^2 \right) \;, \\
 \label{eq:AXMassEstimate}
 m_{A_X}^2 
 &\approx 
 -\frac{3 \, v_X}{\sqrt{2}} \kappa_0 A_0 \left( 1 - 2.3 \, \kappa_0^2 \right) \;.  
\end{align}
To avoid the new scalars becoming tachyonic we find a simple constraint on $A_0$
\begin{equation}
  - \frac{2 \sqrt{2} \, \kappa_0}{ 1 - 2.3 \, \kappa_0^2 } v_X \lesssim  A_0  < 0 \;.
\end{equation}
We checked the approximate formulas for the scalar masses and
found that they are correct up to a few percent, c.f.~Fig.~\ref{Fig:Higgsmasses}.
In particular, $H_X$ receives corrections from finite loop corrections,
which we do not discuss here in detail.
Note that the mass ordering of ${A_X}$ and ${H_X}$ is not fixed, but depends
on $A_0$. This insight helps to separate the $A_X$-funnel from the $H_X$-funnel region.

\section{Sneutrino Dark Matter}
\label{sec:DM}

In this section we discuss the sneutrino as a DM candidate
and focus on (co-)annihilation channels which are unique to our model.
One of them can overcome issues of having a right-handed
sneutrino as a thermally produced DM candidate~\cite{Asaka:2006fs}.
In some sense our approach is similar to the one advertised 
in~\cite{Lee:2007mt},
but we do not introduce a new gauge interaction. 
We only introduce a new complex scalar
which breaks a discrete symmetry and no new vector fields.

This new $X$-scalar splits into a CP-even and a CP-odd scalar after symmetry
breaking such that the new $X$-funnel consists in fact out of two channels,
the $H_X$-funnel and the $A_X$-funnel.
The real and imaginary parts of the lightest sneutrino can then
(co-)annihilate via these channels. 
However, as we will show the
cross section is large enough only for the $A_X$-funnel to realize
the usual freeze-out mechanism.
The pseudoscalar $A_X$ couples predominantly to the heavy
neutrinos, which then decay further into SM particles before big bang nucleosynthesis. 

We focus here mainly on these channels since they are unique to our model.
Other channels like the MSSM Higgs funnels might work as well to realize
sneutrino DM, but these channels have been well studied 
before, e.g., \cite{Fowlie:2012im}.

\subsection{Numerical Calculation of the Particle Spectra}
\label{sec:Spectra}

Before we discuss our results for these channels we describe how 
we determine the particle
spectra. We have assumed GUT-scale boundary
conditions with, in MSSM notation, 
\begin{align}
m_0^2 
 &= 
 \frac{1}{9} m_{\tilde{Q}}^2 = \frac{1}{9} m_{\tilde{D}}^2 = \frac{1}{9} m_{\tilde{U}}^2 = m_{\tilde{L}}^2 = m_{\tilde{E}}^2 = m_{\tilde{N}}^2 = m_{\tilde{S}}^2 = m_{H_u}^2 = m_{H_d}^2 = b_{NS} \;,\\
 M_{1/2} 
 &= 
 \frac{1}{3} M_3 = M_2 = M_1 \;, \\
 A_i 
 &= 
 A_0 Y_i \text{, } A_\lambda = A_0 \lambda \text{, } A_\kappa = \kappa A_0 \;,
\end{align}
inspired by the constrained MSSM.
To avoid the LHC constraints, we have set an arbitrary factor of 3 
for the colored states in order to make them heavy enough. 
The $M_{X}^2$ is fixed at low scale by the tadpole condition as described
in Sec.~\ref{sec:HiggsMasses}. We treat $\tan \beta$, $v_X$, $\kappa$, $\lambda$, 
and $\mu_{NS}$ as free low-scale input parameters and for the sake of simplicity
we choose $\lambda$ and $\mu_{NS}$ to be diagonal. 
We also set $(\mu_{NS})_{22} = 2 (\mu_{NS})_{11}$ and $\lambda_{22} = 2 \, \lambda_{11}$
such that we can consider only one generation of sneutrinos effectively in our scans.
In our model $\kappa$ is an order one parameter 
and we fix it to the reference value $0.4$. 
The reference value for the vev $v_X$ is chosen such that the Xino mass 
is $400$ GeV and our additional fermions and scalars are at the electroweak scale. 

The neutrino Yukawa matrix $Y_\nu$ is in principle free as well,
but we fix it using the tree-level formula
\begin{align}
Y_\nu = \frac{\text{i}}{v_u} U_{\rm PMNS} \sqrt{m_i} \, \Omega \, \left( \sqrt{ M_S^d } \right)^{-1} V_S \, \mu_{NS}\,,
\end{align}
for more details, see Ref.~\cite{Chang:2017qgi}. 
We have used the latest results for normal ordered neutrinos 
from NuFIT~\cite{Esteban:2016qun},  $m_1 = 0$~eV 
and set all CP-violating phases to zero to get real numerical values for $Y_\nu$.
Note that our results depend only very weakly on the details of $Y_\nu$,
but it is important to fix the order of the neutrino Yukawa couplings.

\begin{table}
\begin{center}
\begin{tabular}{ccc}
\toprule
Parameter & $A_X$-funnel & $H_X$-funnel\\
\midrule
$m_0$ & $1\tev$ & $1\tev$\\
$M_{1/2}$ & $1.3\tev$  & $1.3\tev$ \\
$A_0$ &$-250 \leq A_{0}/\gev \leq -50 $ &$-1200<A_{0}/\gev<-350 $\\
$\tan \beta$ & $10$ &$10$ \\
$v_X$& $1000/\sqrt{2}\gev$ &$1000/\sqrt{2}\gev$ \\
$\kappa_0$ & $0.4$ &$0.4$ \\
\midrule
$\lambda_{11} = 0.5 \, \lambda_{22}$  & $1\times 10^{-4}<\lambda_{11}<0.01$  & $1\times 10^{-4}<\lambda_{11}<0.01$  \\
$c$  & \{0.97, 0.99\}  & \{0.97, 0.99\}  \\
\bottomrule
\end{tabular}
\caption{
Parameters used in our numerical scans. We take $\mu_{NS}$ and $\lambda$
to be diagonal matrices. For more details, in particular
on how we determine $\mu_{NS}$ see main text.
}
\label{tab:Parameters}
\end{center}
\end{table}

The calculation of the SUSY spectrum including two-loop corrections 
is taken care of by {\tt SPheno}~\cite{SPheno}. 
The necessary code is generated by {\tt SARAH}~\cite{SARAH}.
In Table~\ref{tab:Parameters}, we give our benchmark parameter ranges for
both funnels which we use in our numerical scans. 

\begin{figure}
 \centering
 \includegraphics[scale=1.2]{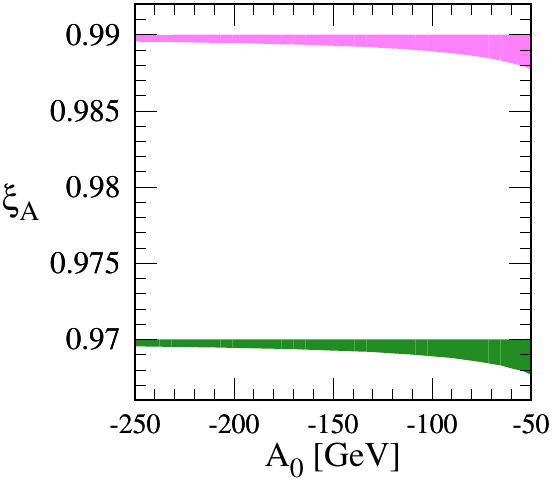}
 \caption{Graphical presentation of our resonance condition
 eq.~\eqref{eq:ResonanceConditionA}. The upper line is for $c = 0.99$
 while the lower line is for $c = 0.97$.
 The ratio $\xi_A$ deviates
 less than  $2.5 \times 10^{-3}$ from the
input value of $c$.
 \label{fig:ResonanceCondition}
 }
\end{figure}

It is numerically very challenging to find the resonance region unless
one understands where to look for it. In our scans we hence implement
an iterative procedure which we describe first for the $A_X$-funnel.
Apart from the fixed parameters in Table~\ref{tab:Parameters} we first
fix $\lambda_{11}$ and $A_0$ to get an estimate for $m_{A_X}$ using
Eq.~\eqref{eq:AXMassEstimate}. Then we choose the parameter $c \in \{0.97, 0.99\}$
which controls how close the data point is to the resonance
\begin{equation}
\label{eq:AFunnelCondition}
m_{\tilde{\nu}^R_1} + m_{\tilde{\nu}^I_1} = c \, m_{A_X} \;.
\end{equation}
We then solve this equation with the formulas~\eqref{Eq:snutree}, 
\eqref{Eq:Rsnucorr}, 
and \eqref{Eq:Isnucorr} from Sec.~\ref{sec:SneutrinoMasses}
to get an initial estimate for $(\mu_{NS})_{11}$:
\begin{align}
\label{eq:AFunnelmuNS}
 (\mu_{NS})_{11} = 
 & {} 
 \frac{c \, m_{A_X}
 \sqrt{ \lambda_{11} ^2 v_X^2 \left(2 \, A_0^2+2 \sqrt{2} \, A_0 \, \kappa \, v_X+ \left(\kappa ^2+2 \, \lambda_{11}^2\right) v_X^2 \right) + 4 \, c^4 \, m_{A_X}^4 - 6 \, c^2 \, \lambda_{11}^2 m_{A_X}^2 v_X^2}
 }{4 \, c^2 \, m_{A_X}^2 - 2 \, \lambda_{11}^2 v_X^2} \nonumber\\
 &{} 
 - \frac{2 \, A_0 \, \lambda_{11}^2 \, v_X^2 + \sqrt{2} \, \kappa \, \lambda_{11}^2 \, v_X^3}{8 \,    c^2 \, m_{A_X}^2 - 4 \, \lambda_{11} ^2 v_X^2} \;.
\end{align}
Given all the initially estimated parameters, 
we then run {\tt SPheno} to calculate a consistent spectrum.
From this calculated spectrum, we take $m_{A_X}$ as reference mass
to calculate again $(\mu_{NS})_{11}$ from Eq.~\eqref{eq:AFunnelmuNS}, 
which we do three times. 
In our final run, we find that the ratio
\begin{equation}
\label{eq:ResonanceConditionA}
\xi_A = \frac{m_{\tilde{\nu}^R_1} + 
m_{\tilde{\nu}^I_1}}{m_{A_X}} \,,
\end{equation}
deviates not more than $2.5 \times 10^{-3}$ from the
input value $c$, 
see also Fig.~\ref{fig:ResonanceCondition}.

To keep the validity of our approximation we have set the constraint
$A_0 < -50$~GeV. For larger and larger $A_0$, the pseudoscalar mass $m_{A_X}$
and hence our sneutrino masses have to become smaller and smaller. This implies
that $(\mu_{NS})_{11}$ should become naively smaller but then the correction term
\begin{equation}
 \frac{m_{\tilde{\nu}^R_1}^2 - m_{\tilde{\nu}^I_1}^2}{m_{\tilde{\nu}^R_1}^2 + m_{\tilde{\nu}^I_1}^2 }
 \approx \frac{\lambda_{11} v_X ( \sqrt{2} \, A_0 - 2 \sqrt{2} (\mu_{NS})_{11} + \kappa \, v_X ) }{4 \, (\mu_{NS})_{11}^2 } \approx \frac{\lambda_{11} \, \kappa \, v_X^2 }{4 \, (\mu_{NS})_{11}^2 } \;,
\end{equation}
is not small anymore since $v_X^2 \gg (\mu_{NS})_{11}^2$ which compensates
the smallness of $\lambda_{11}$. This can also be seen in Fig.~\ref{fig:ResonanceCondition}
where $\xi_A$ deviates more strongly from $c$ for larger $A_0$.

For the $H_X$-funnel we could follow basically the same approach but the relevant
condition to estimate $(\mu_{NS})_{11}$ is modified as
\begin{equation}
 2 \, m_{\chi} = c \, m_{H_X} \,,
\end{equation}
with an estimate for $(\mu_{NS})_{11}$:
\begin{equation}
\label{eq:HFunnelmuNS}
 (\mu_{NS})_{11} = \frac{\lambda_{11} \, v_X}{2 \sqrt{2}} + \frac{1}{2} \sqrt{ \, c^2 \, m_{H_X}^2 - \frac{1}{2} \lambda_{11}  v_X \left(2 \sqrt{2} \, A_0 + 2 \, \kappa \, v_X + \lambda_{11} v_X \right)} \;.
\end{equation}
The resonance parameter in this case reads
\begin{equation}
 \label{eq:ResonanceConditionH}
 \xi_H = \frac{2 \, m_{\chi}}{m_{H_X}} \,,
\end{equation}
which we do not show since we will not use it later explicitly. 
In the $H_X$-funnel parameter set the lightest CP-even sneutrino is always our DM candidate
since it is lighter than the CP-odd sneutrino, i.e.~$m_\chi = m_{\tilde{\nu}_1^R}$ .

With this procedure we have fixed now two sets of particle spectra which can be used in
the further calculations.

\begin{figure}
\begin{center}
\includegraphics[scale=0.45,clip]{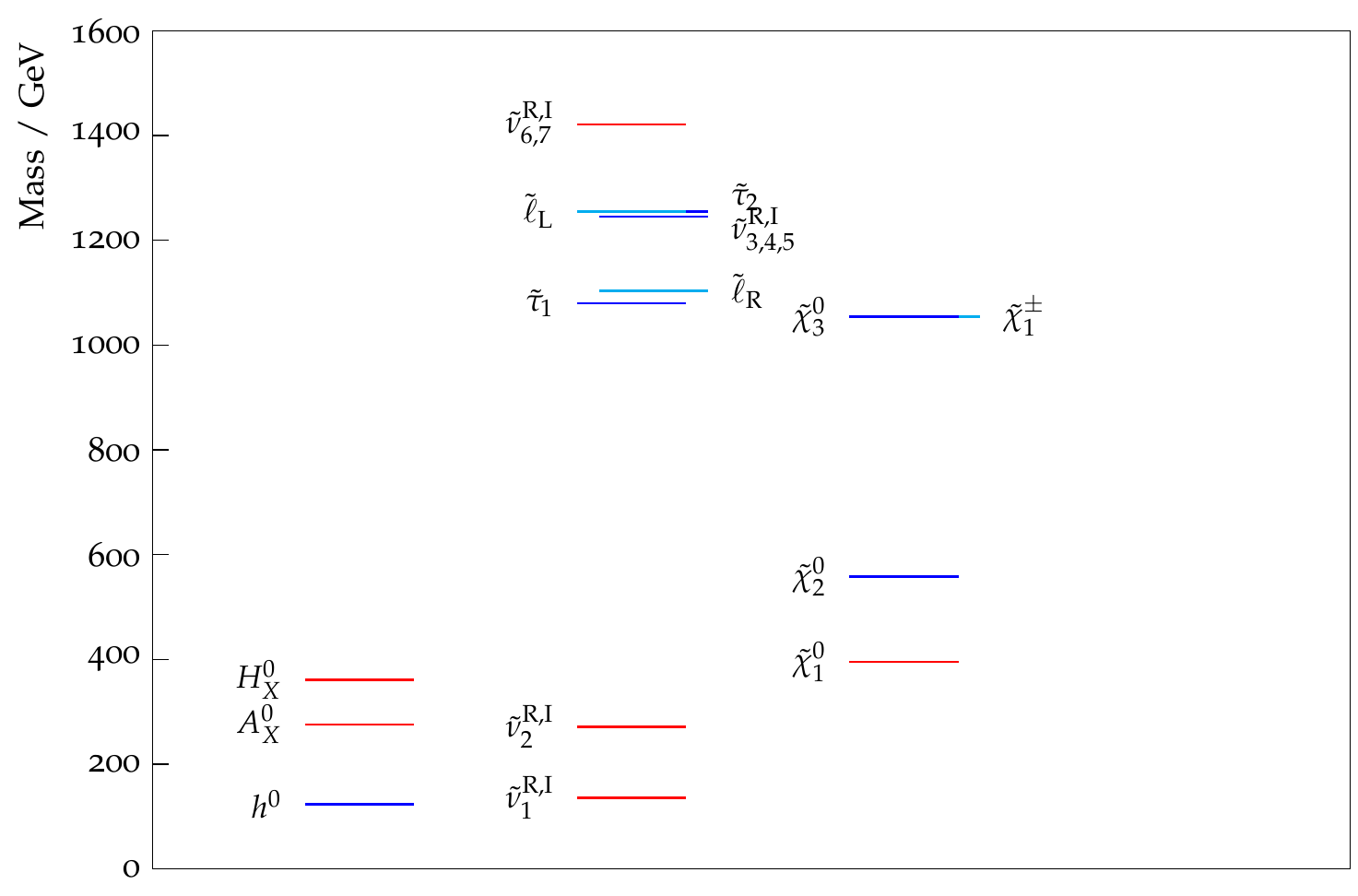}~~
\includegraphics[scale=0.45,clip]{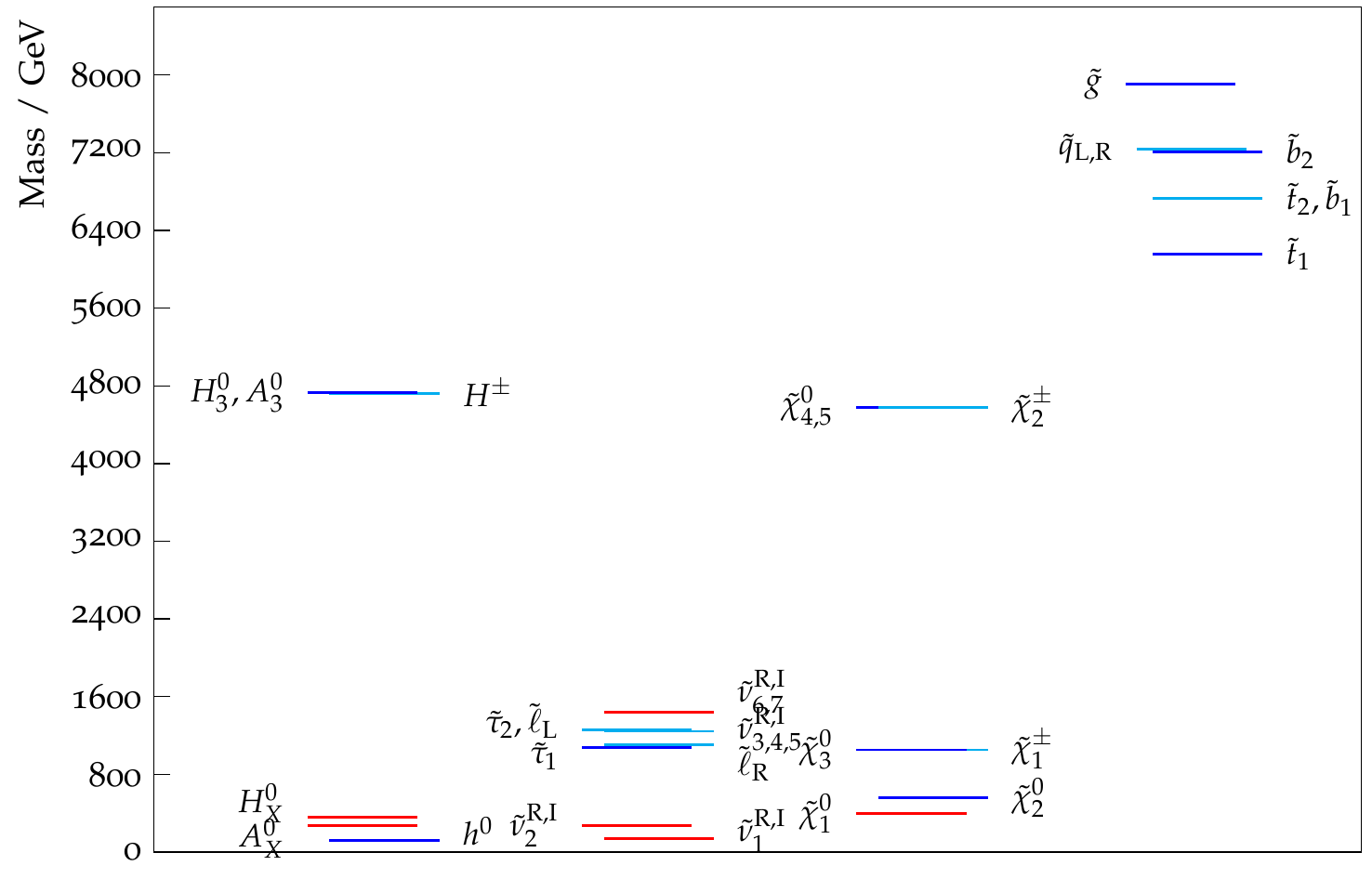}
\caption{
The SUSY mass spectrum of a typical $A_X$-resonance model point. 
The right panel shows the full spectrum and 
the left hand panel is zoomed into the mass scale between 0 to 1.6 $\tev$.
The red lines present the particles beyond the MSSM while blue lines represent the MSSM particles. 
Note that the cyan lines represent degenerate MSSM particles.    
}\label{Fig:mass-spec}
\end{center}
\end{figure}

Last but not least, we present a typical mass spectrum 
of an $A_X$-resonance model point in  Fig.~\ref{Fig:mass-spec}. 
We show its full spectrum in the right panel and  
a zoom-in for smaller masses between 0 to 1.6 $\tev$ in the left panel. 
We use red to represent the particles beyond the MSSM and blue for the MSSM particles. 
Sometimes the masses are very close to each other. 
In this case, we use cyan. 
Note that the MSSM colored particles such as squarks and gluino are heavy enough to evade the LHC constraints. 
Furthermore, the model is very safe from the monojet constraints 
as back-of-the-envelope calculations show that the rates are extremely tiny.

\subsection{(Co-)Annihilation Cross Sections and Relic Density}

\begin{figure}
\begin{center}
\includegraphics[scale=0.7,clip]{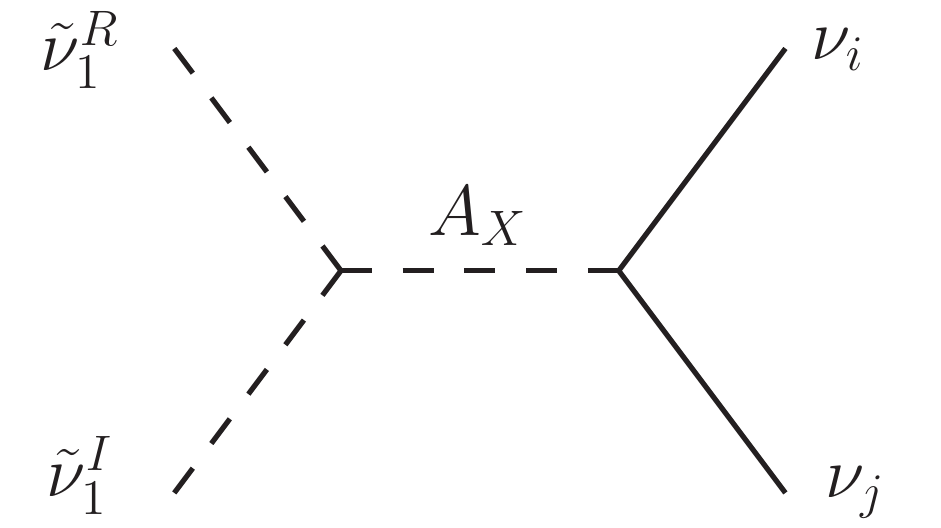}~~
\includegraphics[scale=0.7,clip]{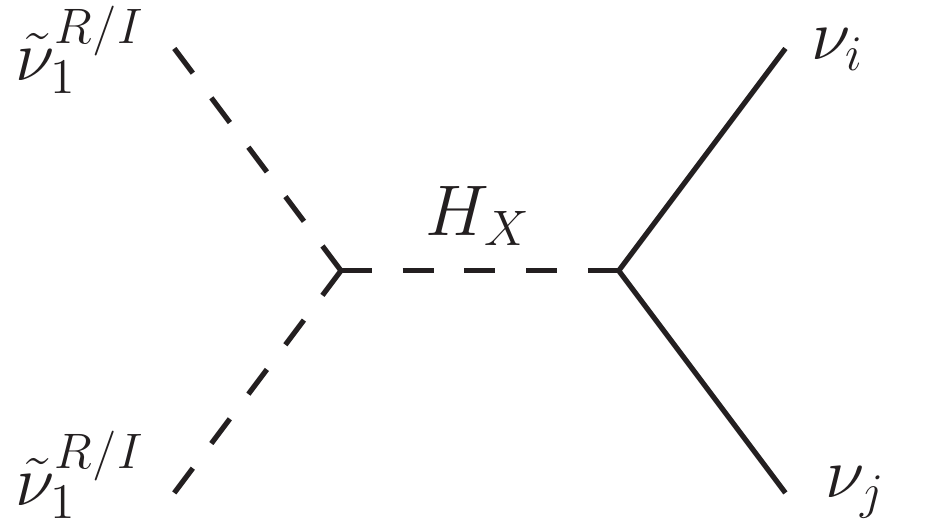}
\caption{The relevant DM annihilation channel for $A_X$-funnel (left) and $H_X$-funnel (right) 
where $i, j = 4,5$. Note that it turns out, that only the $A_X$-funnel is phenomenologically
viable in our setup.
}\label{Fig:DM-annh}
\end{center}
\end{figure}
In our setup, the lightest sneutrino is mostly a maximally mixed combination
of a right-handed sneutrino $\tilde{N}^c$ and the scalar component of the singlet
superfield $\hat{S}$, c.f.~\ref{sec:SneutrinoMasses}. 
The dominant (co-)annihilation channel in the early Universe for them is 
into neutrinos via the $A_X$- or $H_X$-funnel, see Fig.~\ref{Fig:DM-annh}. 
Since we assume CP to be conserved $A_X$ couples to the real and imaginary
part of the sneutrino while $H_X$ to the real part or the imaginary part of the sneutrino.
This implies that the $A_X$-funnel is a co-annihilation channel while 
$H_X$ is an annihilation channel.
In this subsection, we first present the co-annihilation via 
the $A_X$-funnel (scenario A) 
and then discuss why the annihilation via the $H_X$-funnel 
(scenario H) is phenomenologically not viable.

\subsubsection[Scenario A: Sneutrino Co-Annihilation via the AX-Funnel]
{Scenario A: Sneutrino Co-Annihilation via the $\boldsymbol{A_X}$-Funnel}

We begin our discussion with the $A_X$-funnel where $\tilde{\nu}_1^R$ and
$\tilde{\nu}_1^I$ co-annihilate resonantly into neutrinos. 
It is worth to mention that in general resonances can be difficult to handle numerically. 
Furthermore, since our case also has co-annihilation, 
we will start to show complete and approximated formulae for complementarity. 
We have calculated the relic density by using \texttt{MicrOMEGAs} for cross checking 
and we found our relic density computation only differs with \texttt{MicrOMEGAs} $\sim 9\% \mathchar`-10\%$ 
which is properly taken into account by our systematic uncertainties.  
Hence, we perform relic density calculations by ourselves 
using {\tt SPheno}~\cite{SPheno} for the particle spectrum,
as already described in Sec.~\ref{sec:Spectra}.

The co-annihilation cross section into two heavy neutrino states reads
\begin{align}
\sigma (s) = \sum_{i,j = 4}^5
\frac{\left| C_{A \tilde{\nu}\tilde{\nu}} \right|^2 \left| C_{A \nu \nu} \right|^2 \left[ s -\left( |m_{\nu_i}| - |m_{\nu_j}| \right)^2 \right] \mathcal{S}}{8 \pi s \left[ \left( s -m_{A_X}^2 \right)^2 +  m_{A_X}^2 \Gamma_{A_X}^2 \right]}
\sqrt{\frac{\left(  s- m_{\nu_j}^2 \right)^2 - 2 m_{\nu_i}^2 \left(  s + m_{\nu_j}^2 \right) + m_{\nu_i}^4}{\left(  s- m_{\tilde{\nu}_1^R}^2 \right)^2 - 2 m_{\tilde{\nu}_1^I}^2 \left(  s + m_{\tilde{\nu}_1^R}^2 \right) + m_{\tilde{\nu}_1^I}^4}} 
\,,
\label{eq:X-sec-A2} 
\end{align}
where $s = ( p_{ \tilde{\nu}^R_1 } + p_{ \tilde{\nu}^I_1 } )^2 = ( p_{\nu_i} + p_{\nu_j} )^2$ and 
\begin{align}
 \mathcal{S} = \begin{cases}
                1 \text{ for } i \neq j \;, \\
                \tfrac{1}{2} \text{ for } i = j \;.
               \end{cases}
\end{align}
is a symmetry factor. 
We have taken the absolute values of $m_{\nu_i}$ and $m_{\nu_j}$
to emphasize that these are the positive physical masses. 
Here we have three choices for the decay channel, $(i,j)=$ (4,4), (4,5), and (5,5).
The direct annihilation into light active neutrinos is negligibly small. 
Note that the mass difference between the two heavy neutrinos 
is small compared to the mass itself.
$C_{{A} \tilde{\nu}\tilde{\nu}}$ is the coupling of the sneutrinos to $A_X$
\begin{align}
C_{A \tilde{\nu}\tilde{\nu}} 
&= 
- \text{i } 
\Bigg[ 
\kappa \, v_X \lambda_{11} Z_{16}^R Z_{16}^I Z_{23}^A 
- 
\frac{\sqrt{2}}{2} \left( A_\lambda \right)_{11} Z_{16}^R Z_{16}^I Z_{23}^A \nonumber\\
&- 
\frac{\sqrt{2}}{2} \lambda_{11} \left( \mu_{NS} \right)_{11} Z_{14}^R Z_{16}^I Z_{23}^A 
- 
\frac{\sqrt{2}}{2} \lambda_{11} \left( \mu_{NS} \right)_{11} Z_{16}^R Z_{14}^I Z_{23}^A \Bigg]\nonumber\\
&\approx - \text{i } \frac{\lambda_{11}}{2} m_{A_X} \left( \frac{\kappa \, v_X}{m_{A_X}} 
- \frac{1}{\sqrt{2}} \frac{A_0}{m_{A_X}} 
+ \sqrt{2} \frac{\left( \mu_{NS} \right)_{11}}{m_{A_X}} \right) \nonumber\\
&\approx 
- \text{i } \frac{\lambda_{11}}{c} m_{\chi} \left( \frac{c \, \kappa \, v_X}{2 \, m_{\chi}} 
+ \frac{2}{3 \, c} \frac{m_{\chi}}{\kappa \, v_X} 
+ \frac{c}{\sqrt{2}} \right)
\label{eq:CAchichi} \,,
\end{align}
with the mixing matrices of real (imaginary) part of sneutrinos $Z^{R (I)}$
and pseudo-scalars $Z^A$ 
which are $Z^{R (I)}_{14} = -Z^{R (I)}_{16} \approx 1/\sqrt{2}$ 
and $\left| Z^{A}_{23} \right| =1$, respectively. 
For our numerical results we use the full formulas,
but to understand our results it is useful to look at approximate results as well.

Furthermore, $C_{A \nu \nu}$ corresponding to the coupling 
of $A_X$ to the neutrinos is given as 
\begin{align}
C_{A \nu \nu} =
-\frac{1}{\sqrt{2}} \left( \lambda_{11} U_{i6}^V U_{j6}^V + \lambda_{22} U_{i7}^V U_{j7}^V \right) Z_{23}^A  \approx
-\frac{\lambda_{11}}{2 \sqrt{2}} \,,\label{eq:fL}
\end{align}
with the mixing matrices of neutrinos $U^V$ and we have used
that $U_{i6}^V \approx -1/\sqrt{2} \gg U_{i7}^V$ which shows that
$C_{A \nu \nu} $ is of the order of $\lambda_{11}$. 
The total decay width of $A_X$ is dominated by the decay channels 
into heavy neutrinos, $\nu_{4,5}$ or DM 
which can be expressed as 
\begin{align}
\Gamma_{A_X} \Big|_{\rm tot} 
&\simeq
\Gamma_{A_X} \Big|_{A_X \to 2 \nu_4} 
+ 
\Gamma_{A_X} \Big|_{A_X \to 2 \nu_5} 
+ 
\Gamma_{A_X} \Big|_{A_X \to \nu_4 \nu_5} 
+ 
\Gamma_{A_X} \Big|_{A_X \to \tilde{\nu}^R_1 \tilde{\nu}^I_1 }\,, \label{eq:tot-decay}
\end{align}
where we have again neglected the decays into active neutrinos. 
Each term is calculated as follows
\begin{align}
\Gamma_{A_X} \Big|_{A_X \to 2 \nu_4} 
&=
\frac{1}{32 \pi} \lambda_{11}^2 \left( U_{46}^V \right)^4 m_{A_X} \sqrt{1- 4 \left( \frac{m_{\nu_4}}{m_{A_X}} \right)^2} \,, \\
\Gamma_{A_X} \Big|_{A_X \to 2 \nu_5} 
&= 
\Gamma_{A_X} \Big|_{A_X \to 2 \nu_4} (4 \to 5)\,,\\
\Gamma_{A_X} \Big|_{A_X \to \nu_4 \nu_5} 
&= 
\frac{1}{16 \pi} \lambda_{11}^2 \left( U_{46}^V U_{56}^V \right)^2 m_{A_X} 
\left[ 1- \frac{\left( m_{\nu_5} - m_{\nu_4} \right)^2}{m_{A_X}^2} \right]^{3/2}
\sqrt{1- \frac{\left( m_{\nu_5} + m_{\nu_4} \right)^2}{m_{A_X}^2}} \,,\\
\Gamma_{A_X} \Big|_{A_X \to \tilde{\nu}^R_1 \tilde{\nu}^I_1 } 
&= 
\frac{1}{16 \pi} \frac{ |C_{{A} \tilde{\nu}\tilde{\nu}}|^2 }{m_{A_X}} 
\sqrt{1- \frac{\left( m_{\tilde{\nu}_1^R} - m_{\tilde{\nu}_1^I} \right)^2}{m_{A_X}^2}} 
\sqrt{1- \frac{\left( m_{\tilde{\nu}_1^R} + m_{\tilde{\nu}_1^I} \right)^2}{m_{A_X}^2}} \,. 
\end{align}
Using the above equations the total decay width can be approximated by 
\begin{align}
\Gamma_{A_X} \Big|_{\rm tot} \approx 
\frac{\lambda_{11}^2}{64 \pi} 
m_{A_X} \sqrt{1- c^2} 
\times 
\left( 2 + \tilde{C}_{A\tilde{\nu}\tilde{\nu}}^2 \right)
= 
\frac{\lambda_{11}^2}{32 \pi} 
\frac{m_{\chi}}{c} \sqrt{1- c^2} 
\times 
\left( 2 + \tilde{C}_{A\tilde{\nu}\tilde{\nu}}^2 \right)
\,, \label{eq:app-tot-decay}
\end{align} 
where $\tilde{C}_{A\tilde{\nu}\tilde{\nu}}^2$ is 
an $\mathcal{O}(1)$ factor given by 
\begin{align}
\tilde{C}_{A\tilde{\nu}\tilde{\nu}}^2
\equiv
\left( 
\frac{\kappa \, v_X}{m_{A_X}} 
- \frac{1}{\sqrt{2}} \frac{A_0}{m_{A_X}} 
+ \sqrt{2} \frac{\left( \mu_{NS} \right)_{11}}{m_{A_X}} \right)^2
\approx 
\left( 
\frac{c \, \kappa \, v_X}{2 \, m_{\chi}}  
+ \frac{2}{3 \, c} \frac{m_{\chi}}{\kappa \, v_X} 
+ \frac{c}{\sqrt{2}} \right)^2
\,. \label{eq:AFunnelfactor}
\end{align}
Note that this factor depends mildly on $m_\chi$.
We can therefore simplify the thermal averaged cross section 
at the temperature $T$~\cite{Edsjo:1997bg} to 
\begin{align}
\langle \sigma v \rangle_{\rm th}^A 
&= 
\sum_{i,j = 4}^5
\frac{\left| C_{A \tilde{\nu}\tilde{\nu}} \right|^2 \left| C_{A \nu \nu} \right|^2 m_{A_X}^2 \mathcal{S}}{64 \, m_{\chi}^4 \Gamma_{A_X} T} 
\sqrt{\left(1 - \frac{\left( m_{\tilde{\nu}_1^R}+m_{\tilde{\nu}_1^I} \right)^2}{m_{A_X}^2}\right)
\left(1 - \frac{\left( m_{\tilde{\nu}_1^R}  - m_{\tilde{\nu}_1^I} \right)^2}{m_{A_X}^2} \right)^{-1} } \nonumber\\
& \times 
\sqrt{\left( 1 - \frac{\left( m_{\nu_i}+m_{\nu_j} \right)^2}{m_{A_X}^2} \right) 
\left( 1 - \frac{\left( m_{\nu_i} - m_{\nu_j} \right)^2}{m_{A_X}^2} \right)^{3/2}}
\frac{K_1 \left( \frac{m_{A_X}}{T} \right)}{K_2 \left( \frac{m_{\tilde{\nu}_1^R}}{T} \right) K_2 \left( \frac{m_{\tilde{\nu}_1^I}}{T} \right)}
\,,\label{eq:th-ave-A}
\end{align}
with defining $m_\chi \equiv \min \{m_{\tilde{\nu}_1^R}, m_{\tilde{\nu}_1^I} \}$ as the DM mass. 
The functions $K_n (x)$, $n = 1,2$, are the modified Bessel functions of the second kind. 
Here, we have used the narrow width approximation which makes
the $s$-integral easier since the propagator part 
can be replaced by $\pi/(m_{A_X} \Gamma_{A_X}) \delta \left( s -m_{A_X}^2 \right)$.
The total decay width $\Gamma_{A_X}$ is estimated in Eq.~\eqref{eq:tot-decay}. 
This is a complicated formula, but using our previous approximations we find
\begin{align}
\langle \sigma v \rangle_{\rm th}^A 
&\approx 
\sum_{i,j = 4}^5 \mathcal{S}
\frac{\left| C_{A \tilde{\nu}\tilde{\nu}} \right|^2 \left| C_{A \nu \nu} \right|^2}{16 \, m_{\chi}^2 \Gamma_{A_X} T} 
\frac{\sqrt{1 - c^2}}{c^2}
\sqrt{1- \frac{4 \left( \mu_{NS} \right)_{11}^2}{m_{A_X}^2}}
\frac{K_1 \left( \frac{m_{A_X}}{T} \right)}{K_2 \left( \frac{m_{\tilde{\nu}_1^R}}{T} \right) K_2 \left( \frac{m_{\tilde{\nu}_1^I}}{T} \right)} \nonumber\\
&\approx 
\frac{\pi}{4} \frac{\tilde{C}_{A\tilde{\nu}\tilde{\nu}}^2 }{2 + \tilde{C}_{A\tilde{\nu}\tilde{\nu}}^2 } 
\frac{1}{c^3} 
\sqrt{1- \frac{4 \left( \mu_{NS} \right)_{11}^2}{m_{A_X}^2}} 
\; \frac{\lambda_{11}^2}{m_{\chi} T} 
\,\sqrt{\frac{c\, m_\chi}{\pi  \, T}} 
\,\exp \left( - \frac{2 \, (1-c) \, m_\chi}{c \, T} \right)  \nonumber\\
&\approx 
\frac{\pi}{2} 
\frac{\tilde{C}_{A\tilde{\nu}\tilde{\nu}}^2 }{2 + \tilde{C}_{A\tilde{\nu}\tilde{\nu}}^2 } 
\frac{\sqrt{1 - c^2 }}{c^3} 
\; \frac{\lambda_{11}^2}{m_{\chi} T} 
\,{\sqrt{\frac{c \, m_\chi}{\pi \, T}}} 
\,\exp \left( - \frac{2 \, (1-c) \, m_\chi}{c \, T} \right)
\,,\label{eq:th-ave-appr-A}
\end{align}
where we have used the leading order term in the asymptotic expression of the modified Bessel function 
$K_n (x) \approx \sqrt{ \pi / 2 \, x } \, \text{e}^{-x}$~\cite{Srednicki:1988ce}. 
In the second line we have kept the factor with $\left( \mu_{NS} \right)_{11}^2$
as this makes it easier to estimate later how much the $H_X$-funnel is being
more suppressed than this case. 

By using the annihilation cross section formula Eq.~\eqref{eq:th-ave-A}, 
we can numerically solve Eq.~(2) of Ref.~\cite{Griest:1990kh} to obtain the
freeze-out point $x_f = m_\chi/T_f$ 
with the freeze-out temperature $T_f$ 
where we use the entropy table taken from Ref.~\cite{Drees:2015exa}.
We then feed in the freeze-out temperature to calculate the relic density
near the resonance as described in Sec.~V of Ref.~\cite{Griest:1990kh},
\begin{align}
\Omega_{\text{DM}} h^2 = \frac{1.079\times 10^9 \gev^{-1}}{\sqrt{g_*} \, m_{\rm pl} \, J_{x_f} }\,, 
\label{eq:oh2}
\end{align}   
where 
\begin{align} 
J_{x_f} 
&= 
\sum_{i,j = 4}^5 \mathcal{S} \left| C_{A \tilde{\nu}\tilde{\nu}} \right|^2 \left| C_{A \nu \nu} \right|^2  
\frac{\left( m_{\tilde{\nu}_1^R} + m_{\tilde{\nu}_1^I} \right)^2}{4 \, m_{A_X}^5\Gamma_{A_X}}
\sqrt{1-\frac{\left(  \left| m_{\nu_I} \right|  + \left| m_{\nu_J} \right| \right)^2}{m_{A_X}^2}} \notag\\
&
\times
\text{erfc}\left( 
\sqrt{x_f \left(1-\frac{\left( m_{\tilde{\nu}_1^R} + m_{\tilde{\nu}_1^I} \right)^2}{m_{A_X}^2} \right)}
\,
\right)
\left( 1 + \Delta \right)^{3/2} \exp \left( -x_f \Delta \right)
\,,
\end{align}   
with $\Delta \equiv | m_{\tilde{\nu}_1^R} - m_{\tilde{\nu}_1^I} |/m_\chi$. 

Again we can use some approximations to understand
how the relic density scales approximately as
\begin{align}
 J_{x_f} 
 &= 
 \frac{\pi \lambda_{11}^2}{m_{\chi}^2} 
 \frac{\, c^4 \, \tilde{C}_{A \tilde{\nu}\tilde{\nu}}^2}{4 \left( 2 + \tilde{C}_{A \tilde{\nu}\tilde{\nu}}^2 \right)}
\times
\text{erfc}\left( 
\sqrt{x_f \left( 1- c^2 \right)}
\,
\right)
\exp \left( -x_f \Delta \right) 
\,,\label{Eq:Appr-J}
\end{align}
where $\Delta$ can be rewritten by Eq.~\eqref{Eq:Diffcorr} with 
input parameters as 
\begin{align}
\Delta \approx 
\left| 
\frac{\lambda_{11} v_X}{4 \, m_{\chi}^2} 
\left( 
- \frac{8 m_{\chi}^2}{3 \, c^2 \kappa_0 \left( 1 - 2.3 \kappa_0^2 \right) v_X} 
- 2 \sqrt{2} m_{\chi} + \kappa_0 v_X
\right) 
\right|  \,. 
\end{align}
What we can clearly see from this approximation is that
we expect an almost linear behavior of the relic density in the
$(m_\chi, \lambda_{11})$ plane. 
The complementary error function, $\text{erfc}$, is
responsible for the strong dependence of the result on $c$.
For $x_f = 22$ and $c = 0.99$ this piece evaluates to
about 0.35 while for $x_f = 22$ and $c = 0.97$ it gives about 0.11.

\begin{figure}
\centering
\includegraphics[scale=1.2]{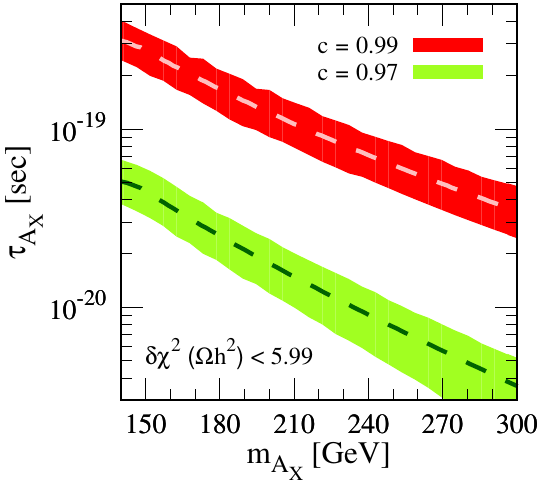} \hspace{0.5cm}
\includegraphics[scale=1.2]{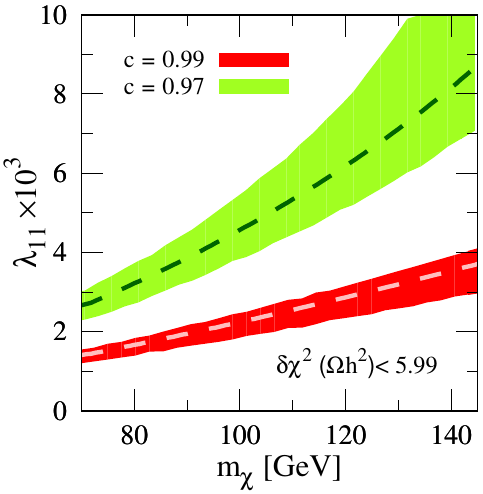}
\caption{Results of our numerical scan where we have applied
a $95\%$ confidence level on the relic density. 
Dashed lines correspond to analytical result applied approximation 
of Eq.~\eqref{Eq:Appr-J} with $x_f = 22$. 
\label{fig:oh2A}}
\end{figure}

In Fig.~\ref{fig:oh2A}, we present the result of our numerical scan where we have
applied a $95\%$ confidence level, namely $\delta\chi^2(\Omega h^2)<5.99$, 
using the PLANCK result~\cite{Ade:2015xua} and including 
conservative $10\%$ theoretical uncertainties.   
The lifetime of $A_X$ is very short which justifies our assumption that the
sneutrinos are in thermal equilibrium with the visible sector before freeze-out.
In the ($m_\chi$, $\lambda_{11}$) plane we see the anticipated approximate linear
relation between $m_\chi$ and $\lambda_{11}$ although, in particular for $c = 0.97$,
there is a non-linear component which is coming mostly from the
$\exp \left( -x_f \Delta \right)$ factor. The dependence of
$\tilde{C}_{A \tilde{\nu}\tilde{\nu}}$ on $m_\chi$ is only subleading compared
to that.

Of particular interest is that the allowed range for $\lambda$ corresponds very
well to the ISS type III scenario discussed in \cite{Chang:2017qgi}.

\subsubsection[Scenario H: Sneutrino Annihilation via the HX-Funnel]{Scenario H: Sneutrino Annihilation via the $\boldsymbol{H_X}$-Funnel}

In the model, there is another potential channel for the DM resonance, i.e.\
the CP-even boson which mostly contains the real component of the $X$ scalar. 
As depicted in the right panel of Fig.~\ref{Fig:DM-annh}, the mediator and the species
of the initial and final states are different from the $A_X$-funnel. Most notably
this channel is a pure annihilation channel. What we will always implicitly assume
is that $\tilde{\nu}^R_1$ is the DM candidate, which is the lighter sneutrino in the
considered parameter range and we set $m_{\tilde{\nu}^R_1} = m_\chi$.

Analogous to the result in Eq.~\eqref{eq:X-sec-A2}  the annihilation cross
section here reads 
\begin{align}
\sigma (s) = \sum_{i,j = 4}^5
\frac{\left| C_{{H} \tilde{\nu}\tilde{\nu}} \right|^2 \left| C_{H \nu \nu} \right|^2 
\left[ s -\left( |m_{\nu_i}| + |m_{\nu_j}| \right)^2 \right] \mathcal{S} }
{8 \pi s \left[ \left( s -m_{H_X}^2 \right)^2 +  m_{H_X}^2 \Gamma_{H_X}^2 \right]}
\sqrt{\frac{\left(  s- m_{\nu_j}^2 \right)^2 - 2 m_{\nu_i}^2 \left(  s + m_{\nu_j}^2 \right)
 + m_{\nu_i}^4}{s \left(  s - 4 m_\chi^2 \right)} } \,,\label{eq:X-sec-H2}
\end{align}
and $C_{{H} \tilde{\nu}\tilde{\nu}}$  and $C_{H \nu \nu}$ are $H_X$-sneutrino-sneutrino 
and $H_X$-$\nu_i$-$\nu_j$ couplings, respectively, 
and they are given as 
\begin{align}
C_{H \tilde{\nu}\tilde{\nu}} 
&= 
\text{i } \Big(
\kappa \, v_X \lambda_{11} Z_{16}^R Z_{16}^I Z_{23}^H 
+ 
\frac{1}{\sqrt{2}} ( A_\lambda )_{11} Z_{16}^R Z_{16}^I Z_{23}^H  \nonumber \\
&  + 
\sqrt{2} \lambda_{11} \left( \mu_{NS} \right)_{11} Z_{14}^R Z_{16}^I Z_{23}^H 
+ 
v_X \lambda_{11}^2 Z_{16}^R Z_{16}^I Z_{23}^H  \Big) \nonumber\\
&\approx 
\text{i }  \frac{\lambda_{11}}{2} m_{H_X} \left( 
\frac{ (\kappa + \lambda_{11})  \, v_X}{m_{H_X}}  + \frac{1}{\sqrt{2}} \frac{A_0}{m_{H_X}} - \sqrt{2}  \frac{\left( \mu_{NS} \right)_{11}}{m_{H_X}} \right)
\,,\label{eq:CHchichi}\\
C_{H \nu \nu} 
&=
- \frac{\text{i}}{\sqrt{2}} \lambda_{11} U_{i6}^V U_{j6}^V Z_{23}^H 
\approx 
- \frac{\text{i}}{2 \sqrt{2}}  \lambda_{11} \,. \label{eq:F45}
\end{align}
The essential difference from the $A_X$-funnel is the factor with 
different mass dependence in the numerator due to the different coupling structure,
i.e.\ here we have $s -\left( |m_{\nu_i}| + |m_{\nu_j}| \right)^2$ while in
the $A_X$-funnel we had $s -\left( |m_{\nu_i}| - |m_{\nu_j}| \right)^2$. 

The partial decay widths for $H_X$ are given by, similar to the $A_X$-funnel case:
\begin{align}
\Gamma_{H_X} \Big|_{H_X \to 2 \nu_4} 
&=
\frac{1}{32 \pi} \lambda_{11}^2 \left( U_{46}^V \right)^4 m_{H_X} 
\left( 1- 4 \left( \frac{m_{\nu_4}}{m_{H_X}} \right)^2 \right)^{3/2} \,, \\
\Gamma_{H_X} \Big|_{H_X \to 2 \nu_5} 
&= 
\Gamma_{H_X} \Big|_{H_X \to 2 \nu_4} (4 \to 5)\,,\\
\Gamma_{H_X} \Big|_{H_X \to \nu_4 \nu_5} 
&= 
\frac{1}{16 \pi} \lambda_{11}^2 \left( U_{46}^V U_{56}^V \right)^2 m_{H_X} 
\sqrt{1- \frac{\left( m_{\nu_5} - m_{\nu_4} \right)^2}{m_{H_X}^2}} 
\left[ 1- \frac{\left( m_{\nu_5} + m_{\nu_4} \right)^2}{m_{H_X}^2} \right]^{3/2} \,,\\
\Gamma_{H_X} \Big|_{H_X \to \tilde{\nu}^R_1 \tilde{\nu}^I_1 } 
&= 
\frac{1}{16 \pi} \frac{ |C_{{H} \tilde{\nu}\tilde{\nu}}|^2}{m_{H_X}} 
\sqrt{1- 4 \left( \frac{m_{\chi}}{m_{H_X}} \right)^2 } \,. 
\end{align}
The thermal averaged cross section for the $H_X$-funnel is given analogous to the $A_X$-funnel as 
\begin{align}
\langle \sigma v \rangle_{\rm th}^H 
&= 
\sum_{i,j = 4}^5 \mathcal{S} \frac{\left| C_{{H} \tilde{\nu}\tilde{\nu}} \right|^2 \left| C_{H \nu \nu} \right|^2 \left[ m_{H_X}^2 - \left( m_{\nu_i}+m_{\nu_j} \right)^2 \right]}{64 \, m_{\chi}^4 \Gamma_{H_X} T} 
\sqrt{1 - 4 \left( \frac{m_{\chi}}{m_{H_X}} \right)^2} \notag\\
& \times 
\sqrt{\left( 1 - \frac{\left( m_{\nu_i}+m_{\nu_j} \right)^2}{m_{H_X}^2} \right) 
\left( 1 - \frac{\left( m_{\nu_i} - m_{\nu_j} \right)^2}{m_{H_X}^2} \right)}
\frac{K_1 \left( \frac{m_{H_X}}{T} \right)}{\left[ K_2 \left( \frac{m_{\chi}}{T} \right) \right]^2}
\,.\label{eq:th-ave-H}
\end{align} 
Similar to Eq.~\eqref{eq:th-ave-appr-A}, the thermal averaged cross section is simplified as 
\begin{align}
\langle \sigma v \rangle_{\rm th}^H \approx
\frac{\pi}{2} \frac{ \tilde{C}_{H \tilde{\nu}\tilde{\nu}}^2 }{2 (1-c^2) + \tilde{C}_{H \tilde{\nu}\tilde{\nu}}^2 } 
\frac{1}{c^3} 
\left( 1- \frac{4 \left( \mu_{NS} \right)_{11}^2}{m_{H_X}^2} \right)^{3/2}
\,\frac{\lambda_{11}^2}{m_{\chi} T} 
\,\sqrt{\frac{c m_\chi}{\pi T}}
\,\exp \left( - \frac{2 (1-c) m_\chi}{c T} \right) 
\,,\label{eq:th-ave-appr-H}
\end{align}
where $\tilde{C}_{H \tilde{\nu}\tilde{\nu}}^2$ is 
\begin{align}
\tilde{C}_{H \tilde{\nu}\tilde{\nu}}^2
\equiv
\left( 
\frac{\left( \kappa + \lambda_{11} \right) v_X}{m_{H_X}} 
+ \frac{1}{\sqrt{2}} \frac{A_0}{m_{H_X}} 
- \sqrt{2} \frac{\left( \mu_{NS} \right)_{11}}{m_{H_X}} 
\right)^2 
\,, \label{eq:HFunnelfactor}
\end{align}
which is also an $\mathcal{O}(1)$ factor. 
Comparing this approximation with Eq.~\eqref{eq:th-ave-appr-A},
we see that the thermal averaged cross section for the $H_X$-funnel 
has an extra suppression by
\begin{equation}
 1- 4 \left( \mu_{NS} \right)_{11}^2/m_{H_X}^2 \sim 1 - c^2 \sim \mathcal{O} (10^{-2}) 
\end{equation}
To compensate for this suppression, we need $\lambda_{11}$ to be larger by a factor
of ten compared to the $A_X$-funnel, roughly speaking.

On the other hand, a larger $\lambda_{11}$ could lead to a 
critical problem of closing
the annihilation channel into heavy neutrinos. If this channel was closed, only the
annihilation into light neutrinos would be allowed, which is heavily suppressed and so the 
annihilation cross section would be too small by far.
To avoid this we need
\begin{align}
 s \approx m_{H_X}^2 > (m_{\nu_i} + m_{\nu_j})^2 \approx 4 \left( \mu_{NS} \right)_{11}^2
 \approx 4 \, m_\chi^2
  + \lambda_{11} \left( 3 \, \kappa \, v_X^2 - \frac{8}{\kappa \, c^2} m_\chi^2 + 2 \sqrt{2} \, m_\chi \, v_X \right)   \;,
\end{align}
where we have used the results from Sec.~\ref{sec:Masses}
and neglected terms of $\mathcal{O}(\lambda^2)$ and $\mathcal{O}(Y_\nu^2)$
and corrections to the masses of the heavy neutrinos.
On the other hand $m_{\chi}^2 = c^2/4 m_{H_X}^2$ and in total
\begin{align}
\lambda_{11} \lesssim \left( \frac{4}{c^2} - 4 \right) m_\chi^2 \left( 3 \, \kappa \, v_X^2 - \frac{8}{\kappa \, c^2} m_\chi^2 + 2 \sqrt{2} \, m_\chi \, v_X \right)^{-1}   \;.
\end{align}
For a typical DM mass $m_\chi=100$~GeV with $c=0.97$, $\kappa=0.4$, 
and $\sqrt{2} \, v_X = 1000$~GeV this means
$\lambda_{11} \lesssim 4.3 \times 10^{-3}$ while from the comparison to the
$A_X$-funnel we expect $\lambda_{11} \sim 2 \times 10^{-2}$.
This rough estimate shows that the $H_X$-funnel is probably not working, which we have
also confirmed numerically. Hence, we conclude that the $H_X$-funnel is not
working within our setup.

\subsection{Dark matter direct and indirect detection}

In this section, we discuss direct and indirect detection prospects
of our DM candidate. We begin with the direct detection 
which is difficult for discovery and then turn to indirect tests 
which are more promising but still far fetched.

\subsubsection{Dark matter direct detection}

\begin{figure}
\begin{center}
\includegraphics[scale=0.7,clip]{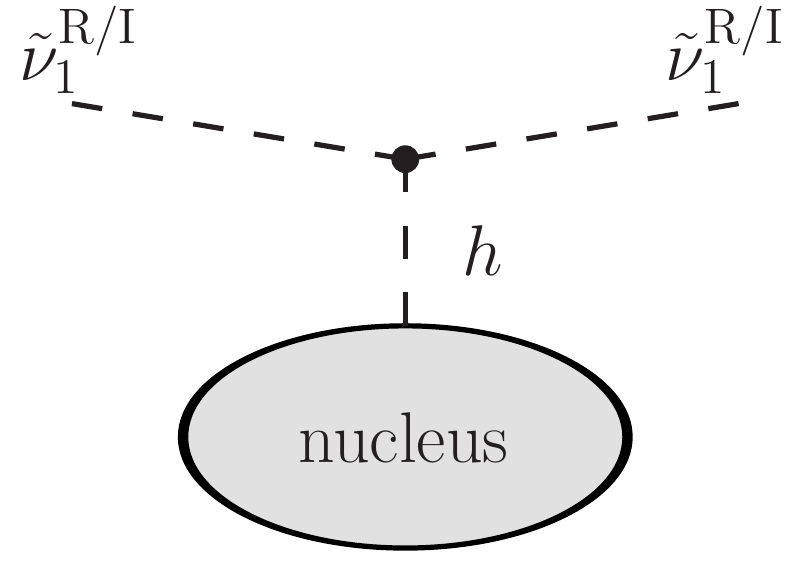}~~
\includegraphics[scale=0.7,clip]{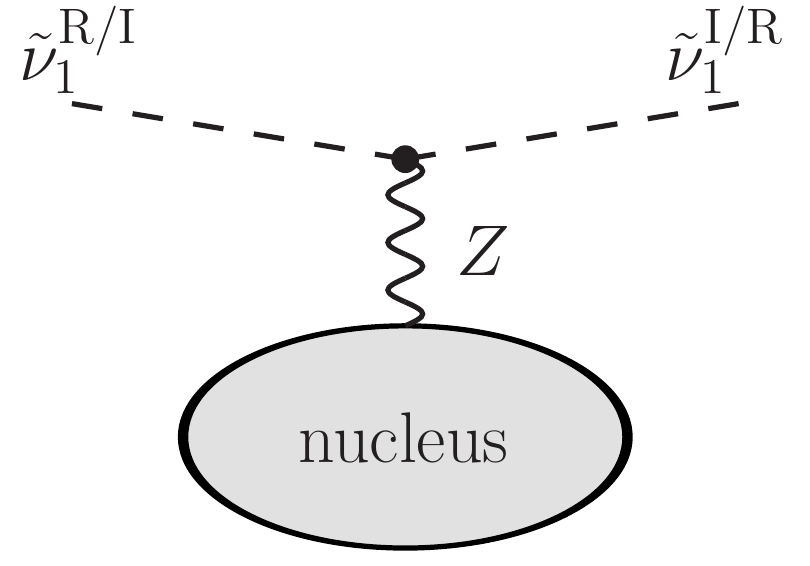}
\caption{The relevant DM direct detection channels. 
}\label{Fig:DM-DD}
\end{center}
\end{figure}

The relevant diagrams for direct detection are shown in Fig.~\ref{Fig:DM-DD}. 
There are two main channels connecting sneutrino DM to the nucleus. 
One is through the SM Higgs boson exchange and another is through the $Z$ boson exchange. 
The relevant effective Lagrangian for DM direct detection is   
\begin{eqnarray}
\mathcal{L}^{\text{Eff.}}_{\text{DD}}\simeq \frac{Y_\nu^2 Y_q M_{\text{SUSY}}}{m_{H}^2} 
(\tilde{\nu}^{R/I}_1 \tilde{\nu}^{R/I}_1)(\bar{q}q) +
\mathcal{C}_Z
(\tilde{\nu}^R_1 \tilde{\nu}^I_1 )(\bar{q} \gamma_5 q)
\,,\label{eq:DD}
\end{eqnarray}
where $M_{\text{SUSY}}$ is of order TeV.
The contribution of the SM Higgs $H$ is larger than the other Higgs bosons
due to either the much larger masses of the other MSSM Higgs bosons or the
small Yukawa couplings to the quarks for $A_X$ and $H_X$. 
The effective coupling $\mathcal{C}_Z$ is 
\begin{align}
\mathcal{C}_Z=\frac{e \, (1+ \tan \theta_W) Y_\nu^2 \, m_q ( g_L^{(q)} - g_R^{(q)} )}{2 \, m_Z^2}
\,, 
\end{align}
with the weak mixing angle $\theta_W$, the quark mass $m_q$, 
and $( g_L^{(q)} - g_R^{(q)} )$ for up-type and down-type quarks is
defined as 
\begin{align}
 g_L^{(d)} - g_R^{(d)} 
 &= 
 \frac{e}{6} \left( \frac{3}{\tan \theta_W} - \tan \theta_W \right) \;,\nonumber \\
 g_L^{(u)} - g_R^{(u)} 
 &= 
 -\frac{e}{2} \left( \frac{1}{\tan \theta_W} + \tan \theta_W \right) \;\nonumber.
\end{align}

We can see that the coefficients in front of the Higgs-exchange 
and the $Z$-exchange terms are roughly of the same order.
On the other hand, the operators themselves give very different 
contributions to direct detection. Due to the presence of 
the $\gamma_5$ matrix in $\bar q \gamma_5 q$, the contribution from
the $Z$-exchange is highly suppressed. 
This is easily understood by considering the nonrelativistic expansion 
of the operator and studying its velocity dependence.

In the non-relativistic limit, the spinors involved in the scattering are
$q = (\xi, \epsilon \, \xi)^T$ and $\bar q = \xi^\dagger (1, \epsilon) \gamma^0$, 
where $\xi$ is the two component Pauli spinor and $\epsilon = O(v/c)$.
The velocity suppression for current DM in the Universe is
$\epsilon \sim  10^{-4} \,\mathchar`- \,10^{-3}$.  We can then do a simple expansion
\begin{align}
  \bar q q 
  &= 
  (\xi^\dagger \;\; \epsilon \, \xi^\dagger) \,
 \left(   \begin{array}{cc}
           1 & 0 \\
           0 & -1  \end{array} \right )\,
 \left( \begin{array}{c}
               \xi \\
            \epsilon \, \xi \end{array} \right ) = (1 - \epsilon^2) \xi^\dagger \xi \;,\\
  \bar q \gamma_5 q &= (\xi^\dagger \;\; \epsilon \, \xi^\dagger) \,
 \left(   \begin{array}{cc}
           1 & 0 \\
           0 & -1  \end{array} \right )\,
 \left(   \begin{array}{cc}
           0 & 1 \\
           1 & 0  \end{array} \right )\,
 \left( \begin{array}{c}
               \xi \\
            \epsilon \, \xi \end{array} \right ) = O(\epsilon^2) \, \xi^\dagger \xi \;.
\end{align}
Therefore, we can easily see that the $Z$-exchange is suppressed by
$O(\epsilon^2) \sim  10^{-8} \,\mathchar`-\, 10^{-6}$ on amplitude level.  We 
conclude that the $Z$-exchange is negligible compared to the Higgs 
exchange contribution and we focus on the Higgs exchange in the following.

The Higgs exchange cross section between DM and a nucleus is given by
\begin{align}
\sigma_{\DM N}(H)= \frac{1}{4\pi} \frac{ m_N^2 }
{(m_\DM +m_N)^2}
\left [ f_p Z+ f_n (A-Z) \right ]^2\,.\label{Eq:DM-scat-H} 
\end{align}
The nucleus parameters, $A$, $Z$, and $m_N$ are 
the mass number, proton number, and the nucleus mass, respectively. 
The effective $\DM \mathchar`-$proton ($f_p$) 
and $\DM \mathchar`-$neutron ($f_n$) couplings for the Higgs channel are 
\begin{align}
f^i_{p,n}= \frac{m_{p,n}}{v} 
\, \left[ \sum_{q=u,d,s} f_{T_q}  \frac{Y_\nu^2 M_{\rm{SUSY}}}
{  m_{H}^2 } +
\sum_{Q=c,b,t} \frac{2}{27} f_{T_G}  \frac{Y_\nu^2  M_{\rm{SUSY}}}
{m_{H}^2}
\right ],
\end{align}
where $v = 246$ GeV.   
The numerical value of nucleon mass matrix elements ($f_{T_q}$ and $f_{T_G}$) 
can be found in \texttt{MicrOMEGAs}~\cite{Belanger:2006is}. 

Using $Y_\nu = 10^{-6}$ and $M_{\rm SUSY} = 1$~TeV 
we find a tiny DM-proton scattering cross section 
$\simeq 10^{-29}$ pb (where $A=Z=1$ in Eq.~\eqref{Eq:DM-scat-H}) 
which is many orders of magnitude below the most stringent current limit
$\sigma_{\DM p}^{\rm{SI}}\sim 5 \times 10^{-11}$ pb for
DM mass at $O(100)$ GeV reported by the
most recent XENON1T \cite{Aprile:2018dbl}.
It is even below the neutrino floor which makes a direct detection
rather difficult. On the other hand, a confirmed direct detection of
DM in the near future would immediately rule out our setup, which is very
attractive.

\subsubsection{Dark matter indirect detection}

Before we discuss indirect detection constraints we want to clarify
that the DM is in fact a two-component dark matter.
From Eq.~\eqref{Eq:Diffcorr} the mass splitting between $\tilde{\nu}^I_1$ and
$\tilde{\nu}^R_1$ is only $O(1)$ GeV. 
Therefore if $\tilde{\nu}^I_1$ is heavier it can decay only into $\tilde{\nu}^R_1$ 
and two active neutrinos via the $A_X$ boson. 
However, the coupling of $A_X$ to the active neutrinos is heavily suppressed and
the phase space is tiny such that the lifetime of  
$\tilde{\nu}^I_1$ is estimated to be much longer than the present age of the Universe. 
The same argument is true if $\tilde{\nu}^R_1$ would be heavier.
Thus, our model is an example of two component DM and since the sneutrinos
are so close in mass we will assume for the sake of simplicity in this section
that they homogeneously form 50\% of DM each.

Since the DM candidate is strongly related to neutrinos,
the most plausible idea is to look for a potential detection 
at IceCube~\cite{Aartsen:2017ulx} via monochromatically produced neutrinos from DM annihilation. 
The DM annihilation $\tilde{\nu}^I_1 \tilde{\nu}^R_1\to \nu_i\nu_j$
via the $A_X$-funnel is the dominant annihilation channel 
at the Galactic center where the DM density is the highest in the Milky Way. 
Here, the indices $i$ and $j$ run from 1 to 5. 
Note that sneutrino DM in this model can annihilate not only into 
heavy neutrinos but also into light active neutrinos but with a large suppression factor.

\begin{figure}
\begin{center}
\includegraphics[scale=0.7,clip]{DM-annih-A}
\includegraphics[scale=0.8,clip]{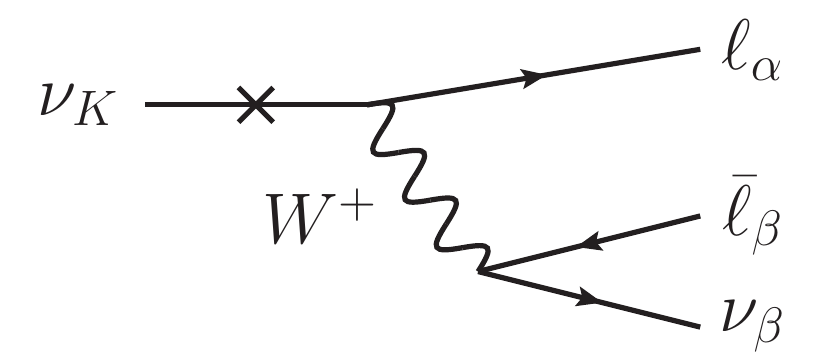}
\caption{(Left) Annihilation channel for DM into a pair of neutrinos,
where $i,j$ runs from 1 to 5. When $i=1\mathchar`-3$ and $j=K=4,5$ corresponds 
to annihilation into one active and one heavy neutrinos.
(Right) The dominant decay of the heavy neutrino
into an active neutrino and charged leptons. Here $K$ is 4 or 5.
}\label{Fig:DM-ID}
\end{center}
\end{figure}

Since the mixing of the singlet fermion(s)
with the active neutrinos is highly suppressed compared to those with heavy neutrinos,
the most obvious channel to produce monochromatic active neutrinos is the annihilation
into one active and one heavy neutrino, followed by subsequent decays
of the heavy neutrino into active neutrinos.
By the analogy from Eq.~(\ref{eq:X-sec-A2}), 
the annihilation cross section of the sneutrino DM into 
one active and one heavy neutrino can be written as,
\begin{align}
\sigma v_{\text{rel}} ( \DM \DM \to \nu_{\rm active} \nu_K )
 = 
\frac{ C_{A \tilde{\nu}\tilde{\nu}}^2 (C_{A \nu \nu}')^2 }
{4 \pi   \left[ \left( s -m_{A_X}^2 \right)^2 +  m_{A_X}^2 \Gamma_{A_X}^2 \right]}
\left(  1 - \frac{ m^2_{\nu_K} }{s} \right )^2 
\,,\label{eq:X-sec-ID}
\end{align}
where $v_{\rm rel}/2 = \sqrt{ 1 - 4 \mchi^2/s }$, the $K = 4$, $5$ represents
the heavy neutrino and active neutrino masses are neglected.
In addition, the coupling $C_{A \nu \nu}'$ is similar to
Eq.~\eqref{eq:fL}
\begin{align}
C_{A \nu \nu}' = -
\frac{1}{\sqrt{2}} \left( \lambda_{11} U_{i6}^V U_{K6}^V 
 + \lambda_{22} U_{i7}^V U_{K7}^V \right ) Z^A_{23}
\,,\label{eq:fLp}
\end{align}
where $i$ runs from $1$ to $3$. 
Since $U_{26}^V \lesssim \mathcal{O} (10^{-6})$
and $U_{37}^V \lesssim \mathcal{O} (10^{-6})$ 
we obtain $\left( C_{A \nu \nu}' \right)^2 \simeq \mathcal{O} (10^{-18})$ after combining
with typical values of $\lambda_{11}$.
Note that due to tiny mixings, the suppression factor 
for direct annihilation into two active neutrinos 
is $\sim\mathcal{O}(10^{-12})$ compared to the annihilation into
one active and one heavy neutrino.

For the DM annihilation into one active and one heavy neutrino 
we hence estimate an annihilation cross section
of $\mathcal{O}(10^{-41})\, \text{cm}^3 \, \text{s}^{-1}$ in our model.
IceCube sets the most stringent limit on DM annihilation cross sections
into monochromatic neutrino lines of
around $2\times 10^{-23}\, \text{cm}^3 \, \text{s}^{-1}$ at 
$m_{\chi} \sim {\cal O}(100)\gev$~\cite{Aartsen:2017ulx} 
using the Navarro-Frenk-White DM profile,
which is much larger than our prediction and a search for monochromatic
neutrino lines from our DM annihilation is not very promising. 
However, there might be some chance to probe the small DM annihilation cross section 
well below $\mathcal{O}(10^{-26})~{\rm cm}^3 {\rm s}^{-1}$ by 
looking for the neutrino line signature boosted by over densities of dark matter spiked around 
a supermassive or an intermediate black hole hosted in spheroidal galaxies~\cite{Arina:2015zoa}.
Naive extrapolation indicates that the quantity energy neutrino times neutrino flux $E \Phi_{\nu_{\rm line}}$ 
is several orders of magnitude below the future sensitivity of IceCube-Gen2.

Nevertheless, the story simply does not end here 
if we consider the heavy neutrino cascade decays into SM particles, 
such as electrons and positrons, and also the bremsstrahlung photons or 
inverse-compton-scattered photons. 
Typical cascade decays of  heavy neutrinos 
into leptons are depicted in the right panel of Fig.~\ref{Fig:DM-ID}. 
For DM annihilation into heavy neutrinos only and subsequent decays into leptons, 
one can take the coupling $C_{A \nu \nu}' \sim \lambda_{11} /2 \sqrt{2}$,
c.f. eq.~\eqref{eq:fL}. 
We found that the annihilation cross section $\langle\sigma v_{\text{rel}}\rangle$ of  
the process shown in Fig.~\ref{Fig:DM-ID} 
can be as large as $\mathcal{O}(10^{-29})\, \text{cm}^3 \, \text{s}^{-1}$. 
This prediction is only a few orders of magnitude below the current
limit of $10^{-26}\, \text{cm}^3 \, \text{s}^{-1}$ from leptophilic DM
channels of the AMS-02 \cite{Cavasonza:2016qem}.
Hence, there may be some chance that 
such charged leptons and secondary photons 
can be probed in future cosmic-ray experiments.

In summary, predictions of our model in both direct and indirect detection
are very safe with the most stringent current limits.
Inversely speaking, once DM direct or indirect detection finds any DM signal, 
our model can be easily excluded by observations.

\section{Summary and Conclusions}
\label{sec:Summary}

Neutrino masses and dark matter are two of the most established
evidences for physics beyond the SM. In a previous work we have
proposed a supersymmetric inverse seesaw model that adds neutrino
masses to the MSSM. The seesaw scale of that model is of the electroweak
scale since it is determined by SUSY breaking parameters. To be more
precise the vev of the scalar component of $\hat{X}$ induces the
lepton number violating terms in the neutrino mass matrix and
the smallness of neutrino masses is then given by some
smallish Yukawa couplings and we can reproduce neutrino oscillation
data perfectly.

In this work we have focused on the novel dark matter
aspects of this model compared to the MSSM.
First of all, the stability of the LSP is guaranteed because of the
unbroken DM parity embedded in our model which acts like matter parity
on the MSSM fields and hence forbids all $R$-parity violating
operators.
The sneutrinos which are odd under this parity thus form a potential
dark matter candidate. In this work the lightest sneutrinos 
which are mainly a linear combination of the scalar components of
the right-handed neutrino $\hat{N}^c$ and singlet $\hat{S}$ superfields
are thermally produced DM particles. 

We have shown that the co-annihilation rate of the CP-even and the
CP-odd sneutrinos can be sizable around the resonance peak of the
pseudoscalar boson $A_X$, the $A_X$-funnel, 
and gives the right amount of DM relic density of the
Universe. The $A_X$ is the pseudoscalar component of the $\hat{X}$
superfield which plays the crucial role to generate the light neutrino
masses as mentioned above. Hence, in our model we have a very close
relationship between the neutrino and the DM sector. Note also that
the current DM population is a mix of CP-even and CP-odd sneutrinos
in our setup.

Furthermore, we have estimated the scattering rate of the sneutrino DM
with nuclei, and found that the dominant contribution comes from the
Higgs-boson exchange.  
Yet, it is many orders of magnitude below the current DM direct detection
limits. Thus, any signals in direct-detection experiments can immediately
rule out our setup. Of course, in that case we could try to embed, for instance,
conventional neutralino DM if in agreement with data.
Similarly, the monochromatic neutrino line signal is a smoking-gun
signature of the model, but the rate is way below the current IceCube limit.  
Nevertheless, the indirect
detection may stand a chance to observe the annihilation of the dark
matter particles into heavy neutrinos followed by their
subsequent cascade decays into charged leptons and photons in cosmic-ray
and gamma-ray telescopes.

Let us again briefly highlight a few important findings of this work
before we conclude:
\begin{enumerate}
\item There is a tiny mass splitting between the real and imaginary 
components of the sneutrino DM. The annihilation via the $A_X$-funnel,
which is a pseudoscalar, has to involve the real and imaginary parts
and is in fact a co-annihilation.

\item The mass of $A_X$ is about two times of the sneutrino DM, such 
that the co-annihilation rate can be sufficiently enhanced so as not to
overclose the Universe.

\item In principal, there is the $H_X$-funnel as well but it
turns out that the annihilation rate in this channel is much smaller than 
in the $A_X$-funnel due to $p$-wave suppression and too small to get the
right relic density.  

\item The scattering cross section of the sneutrino DM with nuclei is
extremely small because of the tiny Yukawa couplings with the ordinary
Higgs boson such that it is well below existing limits.

\item There may be some chance to observe the charged leptons or
secondary photons coming coming from
the co-annihilation of the sneutrino DM in the galactic halo.

\item The most striking feature of our DM is the close link
to the neutrino sector through its couplings to the scalar
$X$-bosons. This is reflected in its phenomenology as well since
DM co-annihilates exclusively into neutrinos.

\item The fermionic component of the $\hat{X}$ superfield, the Xino,
could also be the LSP.  However, its dominant annihilation into
$A_X H_X$ in general has a too large cross section because of the
$\mathcal{O}(1)$ coupling $\kappa$ to give a significant contribution
to the relic density. 
Only when the phase space for this process closes the annihilation cross section
could be sufficiently suppressed.
\end{enumerate}

In this paper we have worked out another striking feature of our
supersymmetric electroweak scale inverse seesaw model, namely the
possible close connection between dark matter and neutrinos.
Hence, apart from the rather technical hierarchy problem
we can solve two of the most outstanding experimental challenges
to the SM, neutrino masses and dark matter. In the future,
we will explore other features of our model related to collider
physics and leptogenesis with hopefully similarly interesting findings.

\section*{Acknowledgements}

 We would like to thank Florian Staub for helping us 
 to implement our model into {\tt SARAH} properly. 
 This research was supported in parts by the Ministry of Science and
 Technology (MoST) of Taiwan under Grant
 No.\ MOST-105-2112-M-007-028-MY3.  J.C.\ was supported by the
 National Research Foundation of Korea (NRF) grant
 No.\ NRF-2016R1E1A1A01943297.


\begin{thebibliography}{99}

\bibitem{silk}
G.~Bertone, D.~Hooper and J.~Silk,
  Phys.\ Rept.\  {\bf 405}, 279 (2005)
  [hep-ph/0404175].
  
  
\bibitem{osci}
Y.~Fukuda {\it et al.} [Super-Kamiokande Collaboration],
  Phys.\ Rev.\ Lett.\  {\bf 81}, 1562 (1998)
  [hep-ex/9807003];
Q.~R.~Ahmad {\it et al.} [SNO Collaboration],
  Phys.\ Rev.\ Lett.\  {\bf 89}, 011301 (2002)
  [nucl-ex/0204008].


\bibitem{Chang:2017qgi} 
  J.~Chang, K.~Cheung, H.~Ishida, C.~T.~Lu, M.~Spinrath and Y.~L.~S.~Tsai,
  JHEP {\bf 1710}, 039 (2017)
  [arXiv:1707.04374 [hep-ph]].


\bibitem{Hagelin:1984wv} 
  J.~S.~Hagelin, G.~L.~Kane and S.~Raby,
  Nucl.\ Phys.\ B {\bf 241}, 638 (1984).


\bibitem{Ibanez:1983kw} 
  L.~E.~Ibanez,
  Phys.\ Lett.\  {\bf 137B}, 160 (1984).


\bibitem{Falk:1994es} 
  T.~Falk, K.~A.~Olive and M.~Srednicki,
  Phys.\ Lett.\ B {\bf 339}, 248 (1994)
  [hep-ph/9409270].


\bibitem{ArkaniHamed:2000bq} 
  N.~Arkani-Hamed, L.~J.~Hall, H.~Murayama, D.~Tucker-Smith and N.~Weiner,
  Phys.\ Rev.\ D {\bf 64}, 115011 (2001)
  [hep-ph/0006312].


\bibitem{Hooper:2004dc} 
  D.~Hooper, J.~March-Russell and S.~M.~West,
  Phys.\ Lett.\ B {\bf 605}, 228 (2005)
  [hep-ph/0410114].


\bibitem{Arina:2007tm} 
  C.~Arina and N.~Fornengo,
  JHEP {\bf 0711}, 029 (2007)
  [arXiv:0709.4477 [hep-ph]].


\bibitem{Arina:2008bb} 
  C.~Arina, F.~Bazzocchi, N.~Fornengo, J.~C.~Romao and J.~W.~F.~Valle,
  Phys.\ Rev.\ Lett.\  {\bf 101}, 161802 (2008)
  [arXiv:0806.3225 [hep-ph]].


\bibitem{Choi:2013fva} 
  K.~Y.~Choi and O.~Seto,
  Phys.\ Rev.\ D {\bf 88}, no. 3, 035005 (2013)
  [arXiv:1305.4322 [hep-ph]].


\bibitem{Asaka:2005cn} 
  T.~Asaka, K.~Ishiwata and T.~Moroi,
  Phys.\ Rev.\ D {\bf 73}, 051301 (2006)
  [hep-ph/0512118].


\bibitem{Asaka:2006fs} 
  T.~Asaka, K.~Ishiwata and T.~Moroi,
  Phys.\ Rev.\ D {\bf 75}, 065001 (2007)
  [hep-ph/0612211].


\bibitem{McDonald:2006if} 
  J.~McDonald,
  JCAP {\bf 0701}, 001 (2007)
  [hep-ph/0609126].


\bibitem{Page:2007sh} 
  V.~Page,
  JHEP {\bf 0704}, 021 (2007)
  [hep-ph/0701266].


\bibitem{Lee:2007mt} 
  H.~S.~Lee, K.~T.~Matchev and S.~Nasri,
  Phys.\ Rev.\ D {\bf 76}, 041302 (2007)
  [hep-ph/0702223 [HEP-PH]].


\bibitem{Cerdeno:2008ep} 
  D.~G.~Cerdeno, C.~Munoz and O.~Seto,
  Phys.\ Rev.\ D {\bf 79}, 023510 (2009)
  [arXiv:0807.3029 [hep-ph]].


\bibitem{DelleRose:2017ukx} 
  L.~Delle Rose, S.~Khalil, S.~J.~D.~King, C.~Marzo, S.~Moretti and C.~S.~Un,
  Phys.\ Rev.\ D {\bf 96} (2017) no.5,  055004
  [arXiv:1702.01808 [hep-ph]];
  L.~Delle Rose, S.~Khalil, S.~J.~D.~King, S.~Kulkarni, C.~Marzo, S.~Moretti and C.~S.~Un,
  [arXiv:1712.05232 [hep-ph]].
  L.~Delle Rose, S.~Khalil, S.~King, J.D., S.~Kulkarni, C.~Marzo, S.~Moretti and C.~S.~Un,
  [arXiv:1804.09470 [hep-ph]].



\bibitem{Farrar:1978xj} 
  G.~R.~Farrar and P.~Fayet,
  Phys.\ Lett.\  {\bf 76B}, 575 (1978).


\bibitem{MatterParity}
  S.~Dimopoulos and H.~Georgi,
  Nucl.\ Phys.\ B {\bf 193} (1981) 150;
  S.~Weinberg,
  Phys.\ Rev.\ D {\bf 26} (1982) 287;
  N.~Sakai and T.~Yanagida,
  Nucl.\ Phys.\ B {\bf 197} (1982) 533;
  S.~Dimopoulos, S.~Raby and F.~Wilczek,
  Phys.\ Lett.\  {\bf 112B} (1982) 133.
  
  
\bibitem{BhupalDev:2012ru} 
  P.~S.~Bhupal Dev, S.~Mondal, B.~Mukhopadhyaya and S.~Roy,
  JHEP {\bf 1209}, 110 (2012)
  [arXiv:1207.6542 [hep-ph]].


\bibitem{Banerjee:2013fga} 
  S.~Banerjee, P.~S.~B.~Dev, S.~Mondal, B.~Mukhopadhyaya and S.~Roy,
  JHEP {\bf 1310}, 221 (2013)
  [arXiv:1306.2143 [hep-ph]].


\bibitem{Guo:2013sna} 
  J.~Guo, Z.~Kang, T.~Li and Y.~Liu,
  JHEP {\bf 1402}, 080 (2014)
  [arXiv:1311.3497 [hep-ph]].


\bibitem{Ghosh:2014pwa} 
  D.~K.~Ghosh, S.~Mondal and I.~Saha,
  JCAP {\bf 1502}, no. 02, 035 (2015)
  [arXiv:1405.0206 [hep-ph]].


\bibitem{Cao:2017cjf} 
  J.~Cao, X.~Guo, Y.~He, L.~Shang and Y.~Yue,
  JHEP {\bf 1710}, 044 (2017)
  [arXiv:1707.09626 [hep-ph]].


\bibitem{Gogoladze:2014vea} 
  I.~Gogoladze, B.~He, A.~Mustafayev, S.~Raza and Q.~Shafi,
  JHEP {\bf 1405}, 078 (2014)
  [arXiv:1401.8251 [hep-ph]].


\bibitem{Kang:2011wb} 
  Z.~Kang, J.~Li, T.~Li, T.~Liu and J.~M.~Yang,
  Eur.\ Phys.\ J.\ C {\bf 76}, no. 5, 270 (2016)
  [arXiv:1102.5644 [hep-ph]].


\bibitem{Chen:2015yuz} 
  S.~L.~Chen and Z.~Kang,
  Phys.\ Lett.\ B {\bf 761}, 296 (2016)
  [arXiv:1512.08780 [hep-ph]].


\bibitem{Khalil:2011tb} 
  S.~Khalil, H.~Okada and T.~Toma,
  JHEP {\bf 1107}, 026 (2011)
  [arXiv:1102.4249 [hep-ph]].


\bibitem{An:2011uq} 
  H.~An, P.~S.~B.~Dev, Y.~Cai and R.~N.~Mohapatra,
  Phys.\ Rev.\ Lett.\  {\bf 108}, 081806 (2012)
  [arXiv:1110.1366 [hep-ph]].


\bibitem{Borah:2012bb} 
  D.~Borah,
  J.\ Mod.\ Phys.\  {\bf 3}, 1097 (2012)
  [arXiv:1204.6587 [hep-ph]].


\bibitem{Abdallah:2017gde} 
  W.~Abdallah and S.~Khalil,
  JCAP {\bf 1704}, no. 04, 016 (2017)
  [arXiv:1701.04436 [hep-ph]].



\bibitem{DeRomeri:2012qd} 
  V.~De Romeri and M.~Hirsch,
  JHEP {\bf 1212}, 106 (2012)
  [arXiv:1209.3891 [hep-ph]].


\bibitem{Frank:2017ohg} 
  M.~Frank and \"{O}.~\"{O}zdal,
  Phys.\ Rev.\ D {\bf 97}, no. 1, 015012 (2018)
  [arXiv:1709.04012 [hep-ph]].


\bibitem{SARAH}
  F.~Staub,
  arXiv:0806.0538 [hep-ph];
  F.~Staub,
  Comput.\ Phys.\ Commun.\  {\bf 185} (2014) 1773
  [arXiv:1309.7223 [hep-ph]].

\bibitem{Fowlie:2012im} 
  A.~Fowlie, M.~Kazana, K.~Kowalska, S.~Munir, L.~Roszkowski, E.~M.~Sessolo, S.~Trojanowski and Y.~L.~S.~Tsai,
  Phys.\ Rev.\ D {\bf 86}, 075010 (2012)
  [arXiv:1206.0264 [hep-ph]].


\bibitem{Esteban:2016qun} 
  I.~Esteban, M.~C.~Gonzalez-Garcia, M.~Maltoni, I.~Martinez-Soler and T.~Schwetz,
  JHEP {\bf 1701}, 087 (2017)
  [arXiv:1611.01514 [hep-ph]].


\bibitem{SPheno}
  W.~Porod,
  Comput.\ Phys.\ Commun.\  {\bf 153} (2003) 275
  [hep-ph/0301101];
  W.~Porod and F.~Staub,
  Comput.\ Phys.\ Commun.\  {\bf 183} (2012) 2458
  [arXiv:1104.1573 [hep-ph]].


\bibitem{Edsjo:1997bg} 
  J.~Edsj\"{o} and P.~Gondolo,
  Phys.\ Rev.\ D {\bf 56}, 1879 (1997)
  [hep-ph/9704361].


\bibitem{Srednicki:1988ce} 
  M.~Srednicki, R.~Watkins and K.~A.~Olive,
  Nucl.\ Phys.\ B {\bf 310}, 693 (1988).



\bibitem{Griest:1990kh} 
  K.~Griest and D.~Seckel,
  Phys.\ Rev.\ D {\bf 43}, 3191 (1991).



\bibitem{Drees:2015exa} 
  M.~Drees, F.~Hajkarim and E.~R.~Schmitz,
  JCAP {\bf 1506}, no. 06, 025 (2015)
  [arXiv:1503.03513 [hep-ph]].


\bibitem{Ade:2015xua} 
  P.~A.~R.~Ade {\it et al.} [Planck Collaboration],
  Astron.\ Astrophys.\  {\bf 594}, A13 (2016)
  [arXiv:1502.01589 [astro-ph.CO]].


\bibitem{Belanger:2006is} 
  G.~Belanger, F.~Boudjema, A.~Pukhov and A.~Semenov,
  Comput.\ Phys.\ Commun.\  {\bf 176}, 367 (2007)
  [hep-ph/0607059].


\bibitem{Aprile:2018dbl} 
  E.~Aprile {\it et al.} [XENON Collaboration],
  [arXiv:1805.12562 [astro-ph.CO]].


\bibitem{Aartsen:2017ulx} 
  M.~G.~Aartsen {\it et al.} [IceCube Collaboration],
  Eur.\ Phys.\ J.\ C {\bf 77}, no. 9, 627 (2017)
  [arXiv:1705.08103 [hep-ex]].

\bibitem{Arina:2015zoa} 
  C.~Arina, S.~Kulkarni and J.~Silk,
  Phys.\ Rev.\ D {\bf 92}, no. 8, 083519 (2015)
  [arXiv:1506.08202 [astro-ph.HE]].



\bibitem{Cavasonza:2016qem} 
  L.~A.~Cavasonza, H.~Gast, M.~Kr\"{a}mer, M.~Pellen and S.~Schael,
  Astrophys.\ J.\  {\bf 839}, no. 1, 36 (2017)
  [arXiv:1612.06634 [hep-ph]].
  
  
\end{thebibliography}
\end{document}